%% file: sp.tex
\documentclass[conference,compsoc]{IEEEtran}

\ifCLASSOPTIONcompsoc
  \usepackage[nocompress]{cite}
\else
  \usepackage{cite}
\fi

\ifCLASSINFOpdf
\else
\fi

\usepackage{tikz}
\usepackage{amsmath}

\usepackage{filecontents}

\input{packages}

\input{macro}

\begin{document}
\title{PromoGuardian: Detecting Promotion Abuse Fraud with Multi-Relation Fused Graph Neural Networks}

\author{\IEEEauthorblockN{
Shaofei Li\IEEEauthorrefmark{2},
Xiao Han\IEEEauthorrefmark{2},
Ziqi Zhang\IEEEauthorrefmark{3},
Minyao Hua\IEEEauthorrefmark{4},
Shuli Gao\IEEEauthorrefmark{4}, \\
Zhenkai Liang\IEEEauthorrefmark{5},
Yao Guo\IEEEauthorrefmark{2},
Xiangqun Chen\IEEEauthorrefmark{2},
Ding Li\IEEEauthorrefmark{2}\IEEEauthorrefmark{1}
}
\IEEEauthorblockA{
    \IEEEauthorrefmark{2}Key Laboratory of High-Confidence Software Technologies (MOE), School of Computer Science, Peking University\\
\IEEEauthorrefmark{3}University of Illinois Urbana-Champaign, 
\IEEEauthorrefmark{4}Meituan, \IEEEauthorrefmark{5}National University of Singapore\\
lishaofei@pku.edu.cn, xhan03@stu.pku.edu.cn, ziqi24@illinois.edu\\ \{huaminyao,gaoshuli\}@meituan.com, liangzk@comp.nus.edu.sg\\
\{yaoguo, cherry, ding\_li\}@pku.edu.cn,\\
 \IEEEauthorrefmark{1}Corresponding Author
}
}

\maketitle

\begin{abstract}
    As e-commerce platforms develop, fraudulent activities are increasingly emerging, posing significant threats to the security and stability of these platforms. \textbf{Promotion abuse} is one of the fastest-growing types of fraud in recent years and is characterized by users exploiting promotional activities to gain financial benefits from the platform. To investigate this issue, we conduct the first study on promotion abuse fraud in e-commerce platforms \Company. We find that promotion abuse fraud is a group-based fraudulent activity with two types of fraudulent activities: Stocking Up and Cashback Abuse.
    Unlike traditional fraudulent activities such as fake reviews, promotion abuse fraud typically involves ordinary customers conducting legitimate transactions and these two types of fraudulent activities are often intertwined. 
    To address this issue, we propose leveraging additional information from the spatial and temporal perspectives to detect promotion abuse fraud. 
    In this paper, we introduce \toolname, a novel multi-relation fused graph neural network that integrates the spatial and temporal information of transaction data into a homogeneous graph to detect promotion abuse fraud. We conduct extensive experiments on real-world data from \Company, and the results demonstrate that our proposed model outperforms state-of-the-art methods in promotion abuse fraud detection, achieving 93.15\% precision, detecting 2.1 to 5.0 times more fraudsters, and preventing 1.5 to 8.8 times more financial losses in production environments.
\end{abstract}

\IEEEpeerreviewmaketitle

\input{tex/intro}

\input{tex/bg.tex}

\input{tex/threatmodel.tex}

\input{tex/study.tex}

\input{tex/approach.tex}

\input{tex/eval}

\input{tex/relatedwork.tex}

\input{tex/discussion.tex}

\input{tex/conclusion.tex}

\bibliographystyle{plain}
\bibliography{references}

\newpage %

\appendices %

\section{Meta-Review}

The following meta-review was prepared by the program committee for the 2026
IEEE Symposium on Security and Privacy (S\&P) as part of the review process as
detailed in the call for papers.

\subsection{Summary}
The paper introduces \toolname, a graph-based model for detecting promotion abuse fraud in e-commerce. \toolname focuses on coordinated behaviors such as shared retail locations or synchronized purchases. The approach fuses multiple relation types into a graph and applies attention with a semi-supervised autoencoder for anomaly detection. The system is evaluated on real-world data, achieves strong performance, and releases both dataset and code.

\subsection{Scientific Contributions}
\begin{itemize}
\item Provides a New Data Set For Public Use

\item Creates a New Tool to Enable Future Science

\item Provides a Valuable Step Forward in an Established Field
\end{itemize}

\subsection{Reasons for Acceptance}
\begin{enumerate}
\item The paper tackles an underexplored but practically important fraud problem with demonstrated real-world impact.

\item Provides a large-scale, real-world dataset and open-source implementation, enabling future research.

\item Shows strong empirical performance and successful deployment that reduced financial losses.

\item Well-motivated design and thorough evaluation across baselines and ablations.
\end{enumerate}

\subsection{Noteworthy Concerns} %
\begin{enumerate} %
\item Scalability: The paper does not quantify the computational or memory overhead of multi-relation fusion (Equation 5). This leaves uncertainty about scalability beyond \Company's infrastructure.

\item Evaluation Gaps: The analysis of robustness is limited. Broader failure cases are not systematically explored, and the adversarial adaptation is only briefly considered.
\end{enumerate}

\end{document}

%% file: packages.tex
\usepackage{cite}
\usepackage{amsmath,amssymb,amsfonts}
\usepackage{graphicx}
\usepackage{textcomp}
\usepackage{xcolor}
\def\BibTeX{{\rm B\kern-.05em{\sc i\kern-.025em b}\kern-.08em
    T\kern-.1667em\lower.7ex\hbox{E}\kern-.125emX}}

\usepackage[all]{nowidow}
\usepackage[linesnumbered,lined,vlined,ruled,commentsnumbered,noend]{algorithm2e}
\colorlet{RED}{red}
\usepackage{listings}
\usepackage{url}

\usepackage{tcolorbox}

\usepackage{xspace}
\usepackage{multirow}
\usepackage{subfloat}
\usepackage{balance}
\usepackage{ctable}
\usepackage{listings}
\usepackage{color}
\usepackage[nobiblatex]{xurl}

\usepackage{multirow}

\usepackage{algpseudocode}
\usepackage{caption}
\usepackage{hyperref}

\usepackage{acronym}
\usepackage{fancybox}
\usepackage{verbatim}
\usepackage[normalem]{ulem}
\usepackage{float}
\usepackage{newfloat}
\usepackage{mdframed}

\usepackage{enumitem}
\usepackage{booktabs}
\usepackage{pifont}

\usepackage{subfigure}
\usepackage{ulem}
\usepackage[flushleft]{threeparttable}
\usepackage{booktabs}
\usepackage{tabularray}

%% file: macro.tex
\definecolor{red}{RGB}{255,0,0}
\definecolor{green}{RGB}{18,220,168}

\usepackage{pifont}

\newcommand{\del}[1]{\textcolor{red}{}}

\newcommand{\Company}{\textsc{Meituan}\xspace}

\newcommand{\toolname}{\textsc{PromoGuardian}\xspace}

\newcommand{\eat}[1]{}

\acrodef{mlaas}[MLaaS]{Machine Learning as a Service}
\acrodef{sdm}[SDMs]{Stateful Defense Models}
\acrodef{qpa}[QPA]{Query Provenance Analysis}
\acrodef{dnn}[DNNs]{Deep Neural Networks}
\acrodef{gnn}[GNNs]{Graph Neural Networks}
\acrodef{sota}[SOTA]{state-of-the-art}
\acrodef{oars}[OARS]{Oracle-guided Adaptive Rejection Sampling}
\acrodef{asr}[ASR]{Attack Success Rate}
\acrodef{pas}[PAS]{Provenance Anomaly Score}
\acrodef{fpr}[FPR]{False Positive Rate}
\acrodef{ttd}[TTD]{Time-To-Detect}
\acrodef{ml}[ML]{Machine Learning}
\acrodef{rnn}[RNN]{Recurrent Neural Network}
\acrodef{lstm}[LSTM]{Long Short-Term Memory}
\acrodef{lbp}[LBP]{Local Binary Pattern}
\acrodef{kg}[KG]{Knowledge Graph}
\acrodef{ppa}[PPA]{Public Promotion Abuse}
\acrodef{tcs}[TCS]{Temporal Cohesion Score}

\definecolor{patriarch}{rgb}{0.5, 0.0, 0.5}
\definecolor{acmblue}{cmyk}{1, 0.1, 0, 0.1}
\definecolor{acmdarkblue}{cmyk}{1, 0.2, 0, 0.15}
\hypersetup{
  colorlinks,
  citecolor=patriarch,
  linkcolor=patriarch,
  urlcolor=acmdarkblue
}

\newlength{\MaxSizeOfLineNumbers}%
\definecolor{keywordcolor}{rgb}{0.8,0.1,0.5}
\definecolor{lightlightgray}{gray}{.96}
\definecolor{lightgray}{gray}{.925}
\definecolor{medlightgray}{gray}{0.7}
\definecolor{medgray}{gray}{0.4}
\definecolor{darkgray}{gray}{0.35}
\definecolor{nearblack}{gray}{0.15}

\newcommand{\distance}{8pt}

%% file: tex/intro.tex
\section{Introduction}
E-commerce platforms play a pivotal role in the modern economy, offering consumers a convenient and efficient way to purchase goods and services while enabling businesses to expand their reach to a global audience. According to a report by Statista~\cite{statista}, global e-commerce sales are projected to reach over \$7 trillion by 2025, highlighting the tremendous growth and significance of these platforms in the current economic landscape.
However, the rapid expansion of e-commerce has also led to a significant rise in fraudulent activities. 
These fraudulent practices not only undermine the financial stability of e-commerce platforms but also disrupt fair competition within the market. A 2024 report from the Association of Certified Fraud Examiners~\cite{acfe} indicates that businesses worldwide lose approximately 5\% of their annual revenue to fraud, which represents a staggering amount when applied to the global e-commerce sector.

\textit{\textbf{Promotion Abuse}} has become a serious problem in the e-commerce and payment industries, particularly in emerging and fast-growing markets~\cite{mrc, fiarni2022detection,bis, unit21,signifyd}. 
To stimulate sales and attract customers, e-commerce platforms frequently offer subsidies such as promotional discounts for customers or cashback rewards for offline dealers based on sales performance. Unfortunately, these promotional strategies are often exploited by fraudsters, resulting in substantial financial losses for the platforms because the investment in promotions is not yielding the intended genuine increase in sales~\cite{PayShield,Paypal,ubereat}.
A RAVELIN survey identifies promotion abuse fraud as the fastest growing type of fraud in the e-commerce sector, with 52\% of marketplaces reporting an increase in such activities~\cite{FraudTrends2021}. At \Company, one of the largest retail e-commerce platforms with more than 470 million users, promotion abuse fraud leads to an estimated annual financial loss of approximately \$10 million. However, no existing research has focused on this type of fraud, and the lack of effective detection methods has made it a significant challenge for e-commerce platforms.

To investigate the promotion abuse fraud in detail, we conducted the first comprehensive study on the existing known cases of promotion abuse fraud at \Company during the second half of 2024. We analyzed 315 reported cases of promotion abuse fraud, which involved over 5,300 fraudsters and 10,000 fraudulent transactions, from the user service team. Through this analysis, we identified two main types of promotion abuse fraud: \textit{Stocking Up} and \textit{Cashback Abuse}. We then find two characteristics of promotion abuse fraud. First, unlike traditional fraud, promotion abuse usually involves ordinary customers who always perform legitimate transactions for their own purposes. Second, these two types of fraudulent activity show different transaction patterns, but are always intertwined. It poses a significant challenge for e-commerce platforms in detecting and preventing these fraudulent activities.

Existing researchers have proposed various methods to detect fraudsters in various scenarios in e-commerce platforms, such as review fraud~\cite{zhang2024dig,shehnepoor2017netspam,rayana2015collective}, click fraud~\cite{yu2023group,li2021large,zhang2025identifying}, and transaction fraud~\cite{zang2023don,li2021happens,shehnepoor2021dfraud3,zhang2022efraudcom,hooi2016fraudar, wang2023removing,10.1145/3589334.3645706}. These studies show that fraudulent activities are usually organized in groups and target the same products or reviews. Consequently, they model the relationships between users, reviews, and products as homogeneous or heterogeneous graphs and employ graph-based algorithms to detect fraudsters, achieving considerable performance.

However, these methods are ineffective in detecting promotion abuse because most of the fraudsters in promotion abuse are ordinary users who also conduct legitimate transactions for their own purposes. They may participate in fraudulent activities in the long term to make a profit, but also conduct legitimate transactions for their daily usage. This occasional participation dilutes the fraudulent behavior, making it harder to distinguish from legitimate transactions. Additionally, the participants in promotion abuse fraud can be organized through social media or offline communication, which makes it difficult to detect if only the user-product relationships are considered.

In our study, we observe that \textit{spatial and temporal coherence can expose the intrinsic relation between transactions, thereby allowing an accurate discovery of abuse fraud}. 
Through our analysis of the temporal and spatial relations between the promotion abuse fraudsters and normal users, we find that the fraudsters exhibit higher cohesion in both spatial and temporal dimensions than normal users.
For example, to obtain subsidies from the platform, fraudsters are organized by leaders and carried out in nearby location using the same shared promotion link. They also have similar consignees to receive the purchased products for stocking up and reselling. 
This process may last for several days or even weeks, during which the fraudulent transactions show high cohesion in these dimensions. Therefore, we can capture this fraudulent transaction pattern by considering the spatial and temporal information of the transactions.

However, leveraging high cohesion in multiple dimensions presents three challenges. 
First, jointly analyzing spatial and temporal relations among users in multiple dimensions is nontrivial, as constructing separate relation graphs for different dimensions fails to capture comprehensive user relationships. To address this, we propose a relation fusion method that constructs a fused homogeneous user graph with fused relational features encoded on the edges.  
Second, fraudsters exhibit diverse behaviors and varying levels of cohesion. Some act as frequent organizers, while others are occasional participants. To prioritize severe behaviors, we employ an attention mechanism in the graph aggregation process, assigning importance weights to user connections to better distinguish anomalous behaviors.  
Third, evolving promotion strategies lead fraudsters to adapt, exacerbating the challenge of limited labeled data (typically less than 10\%). To address this, we propose a semi-supervised learning approach using an autoencoder, enhancing scalability and avoiding reliance on user statistical transaction features, which are susceptible to changes in promotion strategies.

In summary, our main contribution is the proposal of a novel multi-relation fused graph neural networks, \toolname, for group-based promotion abuse fraudsters detection in e-commerce platforms. Unlike existing fraud detection methods that ignore the intrinsic relations between transactions, \toolname captures the high cohesion in spatial and temporal dimensions of fraudulent transactions. Our major contributions are as follows:
\begin{itemize}[noitemsep, topsep=1pt, partopsep=1pt, listparindent=\parindent, leftmargin=*]
    \item We conduct the first study on promotion abuse fraud to analyze its fraud forms and disclose its characteristics. We find that promotion abuse fraudsters exhibit high cohesion in multiple dimensions, including spatial and temporal relations. This cohesion is a fundamental characteristic of fraudulent transactions, but it is often overlooked by existing fraud detection methods.
    \item We propose \toolname, a novel multi-relation fused graph neural network that leverages spatial and temporal relations between transactions to detect fraudsters in group-based promotion abuse. 
    \item We evaluate \toolname on a real-world dataset from \Company, one of the largest e-commerce platforms, and demonstrate its effectiveness in detecting promotion abuse fraudsters compared with five state-of-the-art methods. We also deploy \toolname in the production environment of \Company and evaluate its performance in the reduction of financial losses.l losses.
    \item We publicly release the source code of \toolname and the \ac{ppa} dataset for promotion abuse fraud detection research at \url{https://github.com/0xllssFF/PromoGuardian}. The \ac{ppa} contains anonymized transaction data involving promotions from \Company over a two-week period. This release aims to facilitate future research on fraud detection in e-commerce platforms.
\end{itemize}

%% file: tex/bg.tex
\section{Background and Preliminaries}
\subsection{Fraud Detection in E-commerce Platforms}

Fraud detection is crucial to prevent financial losses and ensure fair competition. 
The detection process typically consists of two phases: offline training and online detection.
In the offline phase, platforms collect historical transaction data and train classification models to identify fraudulent behaviors and manipulate a list of fraudsters. In the online phase, when new orders are placed, platforms refer to this list to determine if the transactions are conducted by fraudsters.  
If so, platforms assess the risk based on manually crafted rules and take appropriate actions, such as approving or blocking the transactions. The platforms' actions depend on the risk level of the transactions, considering factors such as the transaction amount, the number of fraudsters involved, and the frequency of fraudulent transactions.

This paper aims to design effective models to detect fraudsters involved in promotion abuse. Researchers have proposed various graph-based methods to detect fraudsters on e-commerce platforms in different scenarios, such as review fraud~\cite{zang2023don, zhang2024dig,shehnepoor2017netspam,rayana2015collective}, click fraud~\cite{yu2023group,li2021large,zhang2025identifying}, and transaction fraud~\cite{li2021happens,shehnepoor2021dfraud3,zhang2022efraudcom,hooi2016fraudar, wang2023removing,10.1145/3589334.3645706}. These studies highlight that fraudsters often engage in consistent camouflage behaviors to conceal their fraudulent activities~\cite{zang2023don,wang2023removing}. For instance, fraudsters involved in review fraud write fake reviews to promote products and also review other products to conceal their fraudulent activities. To address this camouflage, researchers employ techniques for camouflage identification and removal~\cite{wang2023removing} or carefully select neighbors for aggregation~\cite{zhang2024dig}. However, these techniques are less effective for detecting promotion abuse fraud because fraudsters involved in promotion abuse do not exhibit consistent camouflage behaviors. Often, they are ordinary users unaware of platform rules who occasionally participate in fraudulent activities, possibly as friends or relatives of the organizer. These users conduct normal transactions for their own purposes, resulting in opportunistic camouflage behaviors. This inconsistency makes it challenging to detect fraudsters involved in promotion abuse.

\subsection{Preliminaries} \label{sec:preliminaries}
\noindent \textbf{Relations between Users.} 
Users can be connected through transactional behavior, where relations are established if transactions of two users share the same value in a specific relation within a time window $T$. Formally, a relation $(v_{i}, r_{k}, v_{j})$ exists if transactions of users $v_{i}$ and $v_{j}$ involve purchasing the same products on the same day and share the same value in relation $r_{k}$.

\noindent\textbf{Single-Relation Homogeneous Graph.} 
For each relation $r_{k}$ between users, we can construct an undirected homogeneous graph $\mathcal G_{k} = \{\mathcal V_{k}, \mathcal E_{k}\}$, where $\mathcal V_{k} = \{v_{1}, ..., v_{|\mathcal V_{k}|}\}$ is the set of nodes, $\mathcal E_{k} = \{e_{1}, ..., e_{|\mathcal E_{k}|}\}$ is the edge set. Two nodes $v_{i}$ and $v_{j}$ are connected by an edge if there is a relation $r_{k}$ between the nodes $v_{i}$ and $v_{j}$. 

\noindent\textbf{Multi-Relation Fused Homogeneous Graph.} The graph is an undirected homogeneous graph where the nodes represent the users and the edges represent the fused relation between the users. It is the fusion of a set of single-relation homogeneous graphs. We denote this graph by $\mathcal G = \{\mathcal R, \mathcal V,  \mathcal E, \mathcal M, \mathcal X, \mathcal F, \mathcal Y\}$, where $\mathcal R$ is the set of relations. $\mathcal V = \bigcup_{i=1}^{|\mathcal R|}\mathcal V_{i}$ is the union of the nodes of $|\mathcal R|$ single-relation graphs and $\mathcal E = \bigcup_{i=1}^{|\mathcal R|}\mathcal E_{i}$ is the union of the edge set, which remains only one edge if there are multiple relations between the nodes.
$\mathcal M \in \mathbb{R}^{|\mathcal E| \times |\mathcal R|}$ is the relation map and $m_{ij} \in {\{0, 1\}}^{|\mathcal R|}$ characterizes the relation between two nodes $i$ and $j$, where $m_{ij}(r_{k}) = 1$ if there is a relation $k$ between the nodes $i$ and $j$, and $m_{ij}(r_{k}) = 0$ otherwise. 
$\mathcal X \in {\{1\}}^{|\mathcal V| \times |D_{n}|}$ is the initial node feature set and $D_{n}$ dimensions for each node. 
$\mathcal F \in \mathbb{R}^{|\mathcal E| \times |D_{e}|} $ is the edge feature set and $D_{e}$ dimensions for each edge, which will be introduced in Section~\ref{sec:design:relation}.
$\mathcal Y = \{y_{1}, ..., y_{|\mathcal V|}\}$ is the label set, where $y_{i} \in \{0, -1, 1\}$ with $-1$ denoting the unlabeled nodes, $0$ denoting the normal nodes and $1$ denoting the fraudster nodes.

\noindent \textbf{Problem Statement.} Given a multi-relation fused homogeneous graph $\mathcal G = \{\mathcal R, \mathcal V,  \mathcal E, \mathcal M, \mathcal X, \mathcal F, \mathcal Y\}$, the goal of fraud detection is to predict the label of the nodes in $\mathcal V$ based on the relation between the nodes and the edge features.

%% file: tex/threatmodel.tex
\section{Threat Model}
\label{sec:threatmodel}

In this paper, we focus on the promotion abuse fraud in e-commerce platforms, in which fraudsters exploit the promotion strategies of the platform to make a profit and cause financial losses to the platform. 

\noindent \textbf{Fraudsters.} The fraudsters in promotion abuse fraud are always organized in crowds. They own valid accounts on the platform and they can purchase products from the platform. They have the capability to use multiple devices, create multiple accounts, and organize other fraudsters through social media or offline communication.  
The transactions are trusted and the fraudsters cannot manipulate the transactions. They aim to get a subsidy from the platform or make a profit by reselling at a higher price. 

\noindent \textbf{Platforms.} The platforms frequently introduce promotional incentive activities to attract consumers and increase sales volume. The promotions are well-designed and evaluated by the risk control team. Thus, the fraudsters related to the fraudulent transactions are less than 5\%~\cite{signifyd,Mastercard}. The platform only has information on the transactions. The platforms aim to detect the fraudulent crowds as risky users and punish the leader of the fraudsters. Then, the platforms will limit the subsidy, restrict the promotion, or even block the transactions for the risky users.

%% file: tex/study.tex
\section{Study of Promotion Abuse Fraud}
\label{sec:study}
In this section, we conduct an empirical analysis to demonstrate the impact of promotion abuse fraud on e-commerce platforms and analyze its characteristics. First, we study promotion abuse fraud in \Company during the second half of 2024 to illustrate the scenario and context of such fraudulent activities. Next, we analyze the spatial and temporal characteristics of these fraudulent behaviors using an anonymized public dataset, \ac{ppa}, which contains two weeks of transaction data from \Company's e-commerce platform. Finally, we present a motivating example to illustrate the characteristics of promotion abuse fraud.

\subsection{Promotion Abuse Activities in \Company} \label{sec:study:company}

\noindent \textbf{Businesses of \Company.} This study focuses on promotion abuse fraud within the online retail business of \Company, which involves the buying and selling of goods over the Internet. \Company also operates a logistics network to facilitate the delivery of products to customers. 
To facilitate this process, \Company establishes offline dealers, referred to as \textit{\textbf{headers}}, to organize product sales. 
Each dealer has his offline retail store and manages sales within his respective regions, receives delivered products, and distributes them to customers. They also play a critical role in promoting products and driving sales. The platform offers various promotional activities to stimulate sales, including discounts, coupons and cashback offers for headers. These promotions are designed to attract customers and incentivize headers to increase sales. Many of these promotions are time-sensitive and target specific products or user groups. Over a six-month period, the platform conducts over 300 promotional activities, with 52\% being short-term promotions lasting less than three days and 17\% being long-term promotions lasting more than a month. These dynamic promotional activities are susceptible to exploitation by fraudsters, resulting in significant financial losses for the platform.

\noindent \textbf{Transaction Data.} 
The analysis of promotion abuse fraud is based on transaction data from \Company, which includes user, product, price, quantity, and eight types of relations relevant to fraud detection. These relations are: \textit{order location} (geohash~\cite{8963346} of the order's location), \textit{shared links} (used to share products), \textit{delivery information} (driver and consignee details), \textit{retail store} (offline dealer ID), \textit{group identification} (user group chat ID, which is sourced from the group chat in the \Company's app, not external social media.), \textit{promotion} (promotion ID), \textit{coupon} (coupon type), and \textit{stimulation} (sales strategies for offline dealers). 
Due to the fact that different promotion activities target different products, the quantity and price of the products vary in different promotion activities.
Therefore we do not use these two features in detection of fraudsters and only evaluate financial effects with these two features. 

\noindent \textbf{Label of Fraudsters.}
~\ding{172}~\textit{Reports from the user service team}: The user service team investigates reports originated primarily from drivers and offline dealers. For example, a driver may observe frequent shipments of identical products in abnormal quantities to a specific offline dealer, indicating potential fraudulent behavior.
And offline dealers may report cases where other dealers collude to fabricate transactions and unfairly claim cashback rewards. 
During the second half of 2024, the user service team reported 315 cases involving 5,373 fraudsters. 
~\ding{173}~\textit{Detection results from deployed fraud detection models}: The risk control team develops XGBoost-based models~\cite{chen2016xgboost} to detect malicious transaction behaviors. These models utilize 239 statistical features for each user, including purchasing quantity and transaction amounts across different categories of goods over the past 7, 14, and 30 days.
The models are trained on historical labels and periodically detect fraudsters, with results sampled and manually verified by the policy team.
These two sources are used to maintain and regularly update the fraudster list, which accounts for approximately 1.5\% of the total users.

\noindent \textbf{Fraudulent Activities.}
We conducted an investigation of the 315 reported cases of promotion abuse fraud in the e-commerce platform \Company. We mainly find the following two types of fraudulent behaviors and design rough rules to identify them: 

\noindent \ding{172} \textbf{Stocking up}: 
The platform offers promotional discounts on certain products to attract new customers and stimulate user traffic. These promotional prices are often lower than those on competing platforms due to subsidies. To prevent abuse, the platform limits the purchase quantity per account. However, organized fraudsters exploit these subsidies by purchasing products in large quantities and reselling them at higher prices for profit. They typically target essential goods, such as oil, milk, and beer, which are easy to resell and yield high profits. To avoid detection, fraudsters often recruit ordinary users to place orders and reward them with cash or gifts. These ordinary users also conduct legitimate transactions for personal use, making it difficult to detect. This type of fraud often involves coordinated purchases within specific time frames, as organizers mobilize the fraudsters to purchase the same products simultaneously.
    
\noindent \textbf{Rule 1}: Let $\mathcal{T}$ be the set of all transactions under a promotion. $\mathcal{P}$ is the set of products involved in these transactions.
For each user $u$, the total purchased quantity on certain product $p \in \mathcal{P}$ is defined as:
$
q(u, p) = \sum_{t \in \mathcal{T}_{u,p}} \text{quantity}(t),
$
where $\mathcal{T}_{u,p} \subseteq \mathcal{T}$ is the set of transactions made by $u$ on product $p$. For a group of users $\mathcal{G}$ in the reported cases, the average purchased quantity is defined as:
$
Q(\mathcal{G},p) =\frac{\sum_{u \in \mathcal{G}} q(u, p)}{|\mathcal{G}|} .
$
Stocking up fraud is identified if the average purchase volume on a certain product of the reported cases significantly exceeds normal purchasing behavior. Formally, let $\mu_{n,p}$ and $\sigma_{n,p}$ denote the mean and standard deviation of quantities purchased under the same promotion by normal users. Then, for a predefined parameter $\kappa$, the group $\mathcal{G}$ is flagged as engaging in stocking up fraud if:
\[
\exists p \in \mathcal{P} : Q(\mathcal{G},p) \geq \mu_{n,p} + \kappa\, \sigma_{n,p}.
\]

\noindent \ding{173} \textbf{Cashback Abuse}:  The platform provides cashback incentives to offline dealers to boost sales for specific products or during designated periods, such as an order commission increased by 20\% for the morning market incentive period, targeting fresh products. 
Fraudulent dealers exploit these incentives by recruiting friends, relatives, or incentivizing strangers to repeatedly purchase products. 
It motivates regular users to participate in cashback abuse fraud.
In severe cases, fraudulent headers collude to fabricate transactions and claim cashback rewards. Each header manages a network of users and may share these networks to expand the pool of participants. Typically, fraudulent accounts place one or two low-priced orders per day, often using coupons or benefiting from platform subsidies. Hundreds to thousands of accounts collaborate to fabricate transactions, enabling headers to earn cashback rewards. This type of fraud is covert and can persist for weeks or months.

\noindent \textbf{Rule 2}: Let $\mathcal{H}$ be the headers for a reported group of users $\mathcal{G}$ and $\mathcal{T}_{h}$ be the set of transactions operated by header $h \in \mathcal{H}$.
From the perspective of the platform, the overall income of header $h$ is defined as: $i(h) = c(h) + r(h)$, where $c(h)$ represents the total cashback amount and $r(h)$ represents the regular income. 
The total cashback amount is defined as: 
$   c(h) = \sum_{t \in \mathcal{T}_{c}} \text{cashback}(t),$
where $\mathcal{T}_{c} \subseteq \mathcal{T}_{h}$ is the set of transactions that have cashback incentives.
The regular income is defined as $r(h) = \sigma \sum_{t \in \mathcal{T}_{h}} \text{sale\_revenue}(t)$, where $\sigma$ is the fixed commission rate.
Cashback abuse is identified if $v(h) = \frac{c(h)}{i(h)}$
is significantly high, which means that the header $h$ earns a large proportion of income from cashback.
Formally, let $\mu_{v}$ and $\sigma_{v}$ denote the mean and standard deviation of normal headers. Then, for a predefined parameter $\kappa$, the header $h$ is flagged as engaging in cashback abuse if:
\[
    v(h) \geq \mu_{v} + \kappa\, \sigma_{v}.
\]

These two types of fraudulent activities undermine the platform's purpose, as the subsidies fail to attract genuine traffic and instead become a tool for fraudsters to generate profits. It is also important to note that these two types of fraudulent behaviors are not mutually exclusive.
For instance, a retail store header may organize friends and relatives to stock up on goods while simultaneously earning the cashback incentives of these products. Following the above rules, we set $\kappa=3$ based on the 3-sigma rule for outlier detection~\cite{3sigma} and identify 268 cases involved stocking-up fraud, while 216 were related to cashback abuse, indicating that these two fraudulent behaviors are often intertwined.
Further analysis reveals that 82\% of users involved in these fraudulent activities are ordinary users who also conducted legitimate transactions in the past month. Transactions not associated with promotions are considered legitimate. This finding suggests that most fraudsters are not professional criminals but rather ordinary users who occasionally engage in fraudulent activities. This behavior contrasts significantly with other types of fraud, where groups of fraudsters manipulate ``farmer machines'' to execute fraudulent transactions. 

Therefore, distinguishing fraudulent transactions from legitimate ones is crucial. Existing research predominantly focuses on user-product relationships to construct relation graphs for fraud detection. However, it often overlooks the spatial and temporal information of transactions, which may lead to a decline in detection performance. In the following section, we analyze the spatial and temporal characteristics of transactions in the \ac{ppa} dataset to demonstrate the importance of these features in fraud detection.

\begin{table}[t!]
    \scriptsize
    \centering
    \caption{Statistics information of \ac{ppa}.}
    \resizebox{0.48\textwidth}{!}{\begin{tblr}{
        colspec={X[r, wd = 1cm] c c c c},
        cell{1}{1-5} = {c},
        cell{2-10}{1} = {l},
        cell{2-10}{2-5} = {r},
        hline{1,11} = {1pt},
        vline{2,3,4,5},
        stretch=0
    }
    \SetCell[r = 2, c = 1]{c} \textbf{Dataset} & \SetCell[r = 2, c = 1]{c} \textbf{Users} & \SetCell[r = 1, c = 1]{c} \textbf{Labeled} & \SetCell[r = 2, c = 1]{c} \textbf{Relations} & \SetCell[r = 2, c = 1]{c} \textbf{Edges} \\ 
    & & \SetCell[r = 1, c = 1]{c} \textbf{Fraudsters} & & \\ \hline
    \SetCell[r = 8, c = 1]{c} \textbf{\ac{ppa}} & \SetCell[r = 8, c = 1]{} 5,693,351  & \SetCell[r = 8, c = 1]{} 1.68\% &  \SetCell[r = 1, c = 1]{l} Order Location ($r_1$) & \SetCell[r = 1, c = 1]{} 1,159,257\\
     \textbf{Promotion}&  &  & \SetCell[r = 1, c = 1]{l} Share Link ($r_2$) & \SetCell[r = 1, c = 1]{}  173,029\\
     \textbf{Abuse}&  &  & \SetCell[r = 1, c = 1]{l} Delivery ($r_3$) & \SetCell[r = 1, c = 1]{} 8,824 \\
    &  &  & \SetCell[r = 1, c = 1]{l} Retail Store ($r_4$) & \SetCell[r = 1, c = 1]{} 7,937,295 \\
    &  &  & \SetCell[r = 1, c = 1]{l} Group ID ($r_5$)& \SetCell[r = 1, c = 1]{} 211,150 \\
    &  &  & \SetCell[r = 1, c = 1]{l} Promotion ($r_6$)& \SetCell[r = 1, c = 1]{}  15,826,572\\
    &  &  & \SetCell[r = 1, c = 1]{l} Coupon ($r_7$)& \SetCell[r = 1, c = 1]{} 1,655,738 \\
    &  &  & \SetCell[r = 1, c = 1]{l} Stimulation ($r_8$)& \SetCell[r = 1, c = 1]{}  3,006,006\\

\end{tblr} }
\label{tab:publicdataset}
\end{table}

\subsection{Analysis of Promotion Abuse Fraud}
\label{sec:study:analysis}
To analyze the characteristics of promotion abuse fraud, we conduct a comparative study using an anonymized public dataset, \ac{ppa}, from \Company. This dataset contains a subset of transaction data from online retail businesses for two weeks. We sampled transactions from selected regions to prevent leakage of sensitive commercial statistics, such as the daily active users and the daily transaction volume.
The data collection and anonymization are discussed in Section~\ref{sec:ethics}.
The dataset includes five million users, eight types of relations, and 29 million edges. 
Due to the large scale of the dataset, labeling all users with high confidence is infeasible. Instead, a subset of users has been labeled with high confidence, while the remainder are classified as unknown.
Fraudulent users are labeled as 1, based on the existing black list provided by \Company, accounting for approximately 1.68\% of the total users. 
Normal labels obtained from the whitelist, and additional normal labels expanded using strict heuristic rules. Specifically, we randomly select 1,000 active headers with no record of fraudulent transactions in their retail stores. For each header, up to 100 users who have conducted at least 10 transactions under the header's retail store are included.  These normal users account for 7.12\% of all users. The rest of the users lacking sufficient information are assigned a label of unknown (-1).
Table~\ref{tab:publicdataset} provides the statistics of \ac{ppa}.

\begin{figure}[t!]
    \centering
    \hspace*{-1cm} %
    \includegraphics[width=0.48\textwidth]{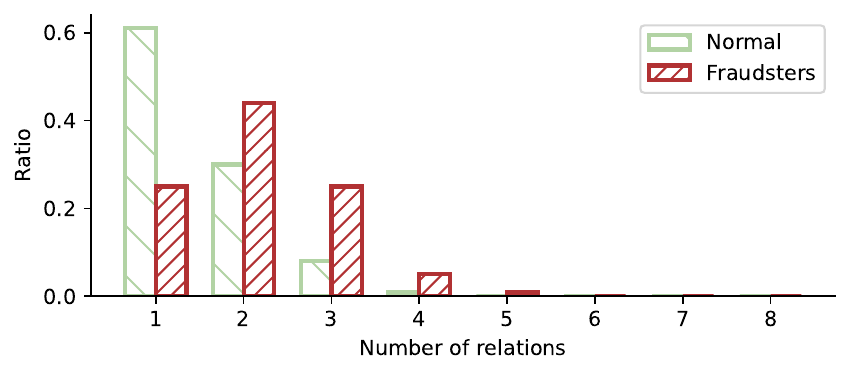}
    \caption{The distribution of the number of relations between users.}
    \label{fig:distribution}

\end{figure}

\noindent \textbf{Spatial Relations.}
To analyze the spatial relations among fraudsters and normal users, we calculate the number of relations among fraudsters and normal users in \ac{ppa} datasets. The number of relations between users is defined as the number of dimensions in which two users have connections. We use this metric to show the spatial relations between users because the fraudsters are well organized and show cohesion in multiple relations, while normal users are more likely not to have these connections. Although these two types of fraudulent activities show different patterns in transaction behaviors, they are both organized in crowds and show high cohesion in multiple relations.
Figure~\ref{fig:distribution} illustrates the distribution of the number of relations between users. Fraudsters exhibit high cohesion across multiple dimensions, with 75\% of their transactions involving more than two relations. In contrast, normal users demonstrate lower cohesion, as 61\% of their transactions involve only one relation. 
Therefore, the fraudsters show high cohesion in multiple relations, while normal users are more likely to have fewer than two relations.

\noindent \textbf{Temporal Relations.}
Considering that these two types of fraudulent activities show different patterns in transaction behaviors in temporal relations, which means that they can occur over short periods or extend over weeks or even months, we propose to use the co-occurrence frequency of users in different spatial relations to quantify the temporal cohesion of fraudsters.
Formally, we propose to leverage the conditional probability of two users $u_{i}$ and $u_{j}$ under spatial relation $r$ to quantify the temporal cohesion. 
The higher $w_{r}^{ij}$, the stronger relationship $r$ between $u_{i}$ and $u_{j}$. We use the transactions in a time window $T$ to calculate $w_{r}^{ij}$ and the definition is shown as follows:
\begin{equation}
w_{r}^{ij} = \frac{freq_{r}(i, j)}{max(freq_{r}(i), freq_{r}(j))}*\beta_{r}^{ij},
\label{eqa:weight}
\end{equation}
\begin{equation}
 \beta_{r}^{ij} = \frac{1}{1 + e^{-\lambda max(freq_{r}(i), freq_{r}(j))}},
\end{equation}
where $freq_{r}(i,j)$ is the frequency of two users appearing together in the dimension $r$ on the same day, $freq_{r}(i)$ and $freq_{r}(j)$ are the frequency of the user $u_{i}$ and $u_{j}$ appearing in the dimension $r$. 
Same-day interactions and the $\beta_{r}^{ij}$ parameter help distinguish genuine connections from coincidental overlaps. 
We calculate $w_{r}^{ij}$ based on same-day interactions because fraudulent users tend to transact within short time frames, often on the same day, while the overall scheme spans longer promotional periods.
The parameter $\beta_{r}^{ij}$ is introduced to distinguish between scenarios where the ratio $ \frac{freq_{r}(i, j)}{max(freq_{r}(i), freq_{r}(j))}$ is identical, but the absolute value of $freq_{r}(i, j)$ and $max(freq_{r}(i), freq_{r}(j))$ differ. 
For example, the situation where $freq_{r}(i, j) = 8$ and $max(freq_{r}(i), freq_{r}(j)) = 10$ differs from the scenario where $freq_{r}(i, j) = 80$ and $max(freq_{r}(i), freq_{r}(j)) = 100$. 
In the latter case, $w_{r}^{ij}$ should be higher because they show stronger cohesion in the dimension $r$.
The hyperparameter $\lambda$ controls the adjustment of the weight, with a default value of $\lambda = 1$.
In this way, $w_{r}^{ij}$ can represent the strength of the relationship between two users in the dimension $r$, no matter whether the fraudulent activities are conducted in a short time or over a long period.

To demonstrate the effectiveness of $w^{ij}_{r}$ in characterizing the temporal cohesion of fraudsters, 
we propose the use of the \ac{tcs}, $s_{c}^{r}$, to evaluate the relative density of intra-group connections compared to connections with external users. Inspired by the concept of modularity in community detection~\cite{newman2004finding,wang2023removing}, we define \ac{tcs} as follows:
\begin{equation}
    s_{c}^{r} = \frac{\sum_{u_{i} \in c} \sum_{u_{j} \in c} w_{r}^{ij}}{\sum_{u_{i} \in c} \sum_{u_{j} \in V} w_{r}^{ij}}
\end{equation}
Here, $c$ denotes the group of users, $V$ represents the set of all the users, $w_{r}^{ij}$ indicates the \ac{tcs} between user $u_{i}$ and $u_{j}$ in the relation $r$. A higher $s_{r}^{c}$ signifies the stronger temporal cohesion for the relation $r$ within the group $c$. 

To investigate the temporal cohesion of different groups in various spatial relations, we calculate the \ac{tcs} for normal and fraudulent groups in different relations and perform significance testing on the results across different regions. 
Specifically, for each region, we select five fraudulent groups and five normal groups. The selection of a subset, rather than all groups, helps to reduce computational cost while maintaining the representativeness of the average scenario.
Users were then categorized into two groups: $c_{n}$, representing normal users, and $c_{f}$, representing fraudsters, with $V = c_{n} \cup c_{f}$. 
For $c_{n}$, we extracted a subgraph of normal users, denoted as $G_{n}$ and for $c_{f}$, we extract the subgraph of the fraudsters, denoted as $G_{f}$. Using the $G_{n}$ and $G_{f}$, we can calculate $s_{c}^{r}$ for both the fraudsters and normal users in the relation $r$. 
We repeat the calculation of \ac{tcs} in 20 regions and then for each relation, we conduct a two-sample t-test~\cite{ttest} to assess the null hypothesis: ``there is no difference in \ac{tcs} between fraudulent and normal groups''. The average \ac{tcs} of $G_{n}$ and $G_{f}$ for each relation is presented in Table~\ref{tab:temporalscore}. All the p-values are less than 0.05, demonstrating that the observed differences in \ac{tcs} are statistically significant across all the relations~\cite{ttest}. 
Fraudsters and normal users show similar \ac{tcs} in $r_{4}$, which is because $r_{4}$ indicates the retail store where the transactions occur. Both fraudsters and normal users tend to shop at the same retail stores, leading to similar \ac{tcs} values.
Overall, this suggests that fraudsters exhibit stronger temporal cohesion in different spatial relations, indicating that they frequently co-occur in various spatial relations.

\begin{table}[t!]
    \scriptsize
    \centering
    \caption{\ac{tcs} of $c_{n}$ and $c_{f}$ in different relations. Higher \ac{tcs} indicates stronger temporal cohesion in certain spatial relations. The p-values are obtained from two-sample t-test to assess the null hypothesis: ``there is no difference in \ac{tcs} between fraudulent and normal groups in $r_{i}$''.}
    \resizebox{0.48\textwidth}{!}{\begin{tblr}{
        cell{1-2}{1-8} = {c},
        cell{3-6}{1,5} = {c},
        cell{3-6}{2-4} = {r},
        cell{3-6}{6-8} = {r},
        hline{1,2,7} = {1pt},
        vline{2,3,4,5,6,7,8},
        stretch=0
    }
    \SetCell[r = 2, c = 1]{c}  \textbf{Relation} &  \SetCell[r = 1, c = 3]{c} \textbf{\ac{tcs}} & & & \SetCell[r = 2, c = 1]{c}  \textbf{Relation} &  \SetCell[r = 1, c = 3]{c} \textbf{\ac{tcs}} & \\
     & \SetCell[r = 1, c = 1]{c} \textbf{$G_{n}$}  & \SetCell[r = 1, c = 1]{c} \textbf{$G_{f}$} & \textbf{p-value} & & \SetCell[r = 1, c = 1]{c} \textbf{$G_{n}$}  & \SetCell[r = 1, c = 1]{c} \textbf{$G_{f}$} & \textbf{p-value} \\\hline
    $r_1$ & 0.48 & 0.69 & $\textless$ 0.01 & $r_5$ & 0.77 & 0.92 & $\textless$ 0.01\\
    $r_2$ & 0.55 & 0.83 & $\textless$ 0.01 & $r_6$ & 0.32 & 0.89 & $\textless$ 0.01\\
    $r_3$ & 0.68 & 0.81 & $\textless$ 0.01 & $r_7$ & 0.54 & 0.87 & $\textless$ 0.01\\
    $r_4$ & 0.85 & 0.90 & 0.027 & $r_8$ & 0.52 & 0.82 & $\textless$ 0.01\\

\end{tblr}}
\label{tab:temporalscore}
\end{table}

\begin{tcolorbox}[title=Conclusion, left=2pt, right=2pt, top=2pt, bottom=2pt]
    The results demonstrate that the fraudsters in promotion abuse fraud show high cohesion patterns in spatial and temporal relations.
    For spatial relations, cohesion is reflected in the higher dimension of interactions between them.
    For temporal relations, the fraudsters in promotion abuse fraud show high co-occurrence frequency on different spatial relations.
    The results motivate our insight in detecting promotion abuse fraud by considering the spatial and temporal relations between transactions.
\end{tcolorbox}

\subsection{Motivating Example}

\begin{figure}[!t]
    \centering
    \includegraphics[width=0.45\textwidth]{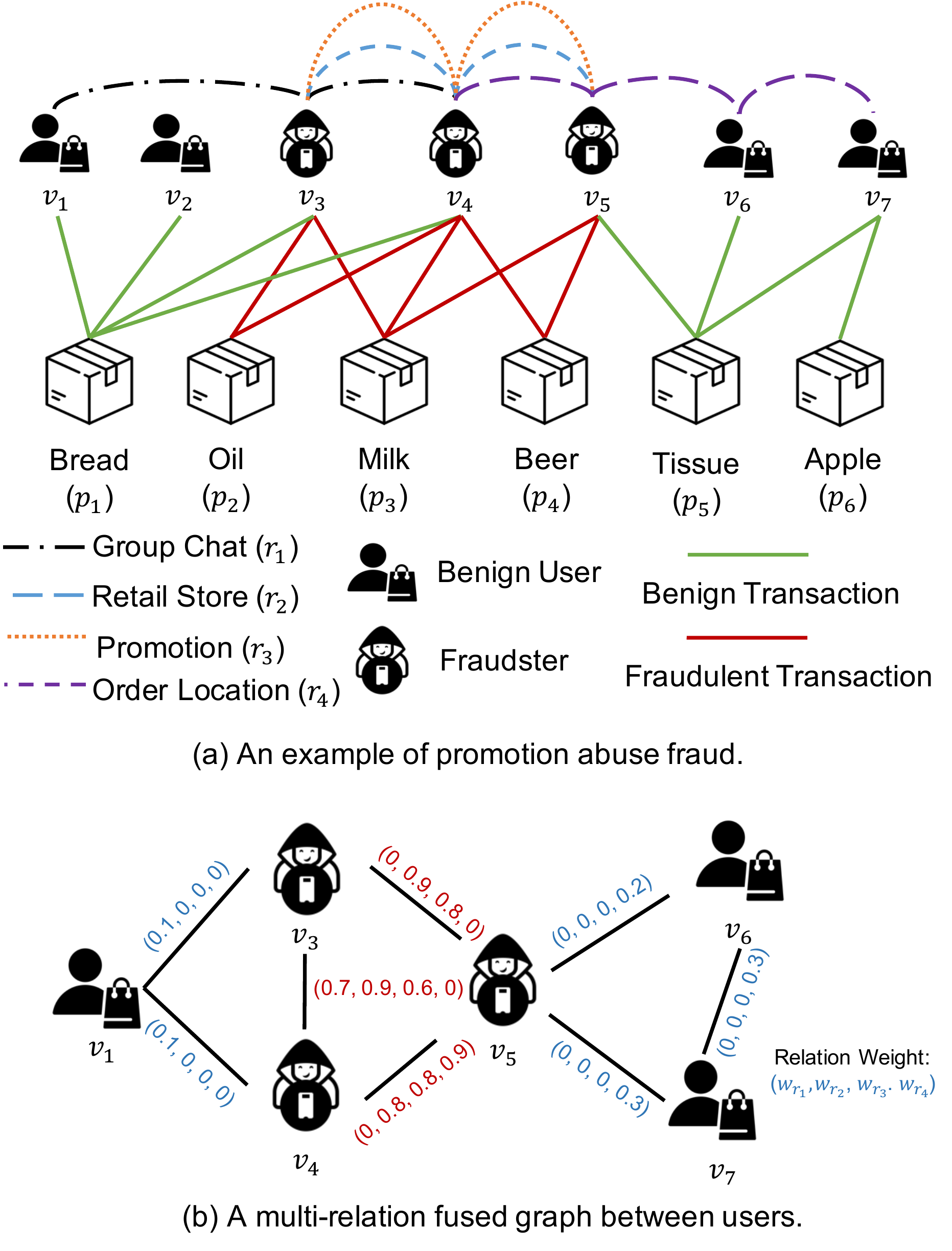}
    \caption{(a) Users and transactions on the e-commerce platform. Users connected with certain products indicate that they purchase this product. Users connected by the same line type indicate that their transactions share the same value in that relation. (b) Users establish connections when their transactions involve the same product and share the same relation value. User $v_{2}$ does not have any relation with other users and is therefore not shown in the fused graph.}
    \label{fig:motivation}
\end{figure}

To illustrate the high cohesion in multiple relations of promotion abuse fraud, we present a motivating example in Figure~\ref{fig:motivation}. Figure~\ref{fig:motivation}(a) depicts a scenario derived from \Company, where the fraudsters ($v_{3}, v_{4}, v_{5}$) are organized into groups to exploit promotional discounts on highly circulated products ($p_{2}, p_{3}, p_{4}$), such as oil, milk, and beer. These products are essential for daily life, have high prices, and are easy to resell.
There are four types of relations between the transactions of users: Group Chat ($r_{1}$), Retail Store ($r_{2}$), Promotion ($r_{3}$), and Order Location ($r_{4}$).
These fraudsters purchase products at discounted prices and resell them for profit. Notably, they may also engage in normal transactions for personal use, exhibiting diverse behavior patterns. 
In this example, $v_{4}$ is an offline dealer that owns a retail store.
$v_{3}$ is organized by $v_{4}$ through online communication, leading them to purchase the same promotional products ($p_{2}, p_{3}$) under the same promotion in the same chat room and have the items delivered to $v_{4}$'s retail store. This results in cohesion across Group Chat ($r_{1}$), Retail Store ($r_{2}$), and Promotion ($r_{3}$) relations. 
Similarly, $v_{5}$ is organized by $v_{4}$ offline, purchasing the same promotional products ($p_{3}, p_{4}$) to the same retail store in the same geohash zone, creating cohesion in a Retail Store ($r_{2}$), Promotion ($r_{3}$) and Order Location ($r_{4}$) relations. This reflects strong spatial cohesion.
The fraudulent transactions span five days, during which the fraudsters exhibit high co-occurrence frequency across these dimensions, demonstrating temporal cohesion. In contrast, benign users also purchase popular products ($p_{1}$ and $p_{5}$) for personal use, but their transactions show lower cohesion in spatial and temporal relations.
This example highlights that high cohesion across multiple relations is a fundamental characteristic of fraudulent transactions. 

Based on this observation, we can model the relations between users and products as a multi-relation fused homogeneous graph, shown in Figure~\ref{fig:motivation}(b). The nodes represent the users and the edges represent the relations between users. For example, user $v_{3}$ and user $v_{4}$ have relations in $r_{1}, r_{2}$ and $r_{3}$ because they purchase the same products and the transactions are consistent in the chat room, promotion and retail store. Each edge is assigned a concatenated relation weight $(w_{r_{1}}, w_{r_{2}}, w_{r_{3}}, w_{r_{4}})$, fused by different spatial and temporal information. The value of the relation weight is determined by the strength of the relation between users, calculated by the co-occurrence frequency of two users in the same relation. We will introduce the details of the multi-relation fused homogeneous graph in Section~\ref{sec:design:relation}. The group-based fraudsters $(v_{3}, v_{4}, v_{5})$ show strong cohesion in multiple relations, labeled in red. In contrast, benign users $(v_{1}, v_{6}, v_{7})$ can also be connected in the same graph but show weak cohesion in multiple relations, labeled in blue.

Traditional group-based fraud detection methods cannot detect it when they treat each relationship individually or assign equal importance to all relationships. For example, considering only user-product relationships, fraudsters and legitimate users may appear densely connected, making it challenging to distinguish between them. This can result in false positives, such as incorrectly identifying $v_{1}, v_{6}, v_{7}$.
Some existing methods attempt to detect fraud by identifying and removing consistent camouflage behaviors exhibited by fraudsters. However, in promotion abuse fraud, fraudsters do not exhibit consistent camouflage patterns. Thus, it is crucial to model user relationships across multiple dimensions and develop a novel fraud detection approach that captures the high cohesion inherent in these relational structures of fraudulent transactions.

%% file: tex/approach.tex
\begin{figure*}[!t]
    \centering
    \includegraphics[width=\textwidth]{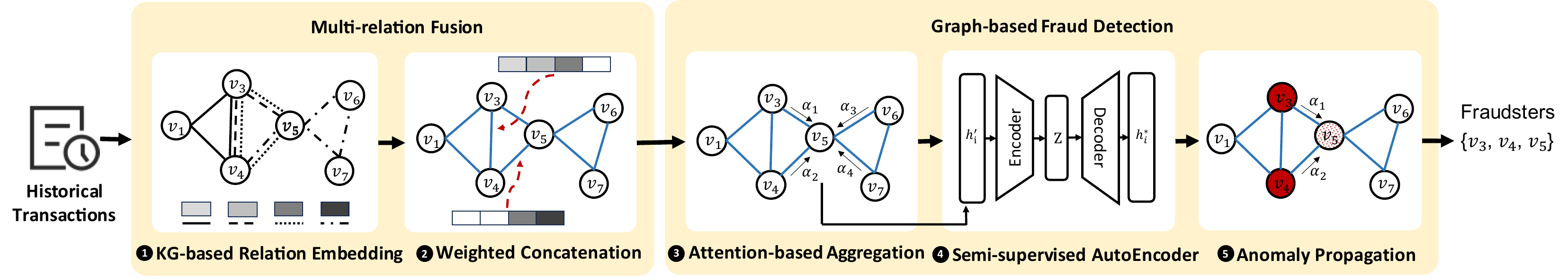}
    \caption{Workflow of \toolname.}
    \label{fig:workflow}
\end{figure*}

\section{Design}\label{sec:design}
In this section, we present the design of \toolname for detecting promotion abuse fraud. We first introduce the multi-relation fusion method to represent the relationship between users in multiple dimensions. Then we present the multi-relation fused homogeneous graph with attention mechanism to conjointly analyze the transaction behavior in multiple dimensions. Finally, we propose a semi-supervised learning method with an autoencoder to enhance the scalability in detecting fraudsters in the real-world production environment. The workflow of \toolname is illustrated in Figure~\ref{fig:workflow}.

\subsection{Multi-relation Fusion}\label{sec:design:relation}

Spatial and temporal relations are crucial for detecting promotion abuse fraud. Relying solely on individual relations may be insufficient to effectively capture fraudulent activities and differentiate them from normal transactions. Therefore, it is essential to explore the relationships between users across multiple dimensions.
To address this challenge, we propose a multi-relation fusion method that integrates spatial and temporal information to represent the complex relationships between users. This method comprises two key steps: relation embedding and weighted concatenation. The relation embedding step utilizes \ac{kg} embeddings to capture intricate user relationships and generate embeddings for each relation. Then the weighted concatenation of these embeddings is performed to incorporate spatial and temporal information, as well as to quantify the strength of user relationships across different dimensions.

\noindent \textbf{\ac{kg}-based Relation Embedding.} 
To initialize the relation embedding, we leverage \ac{kg}~\cite{hogan2021knowledge,10.1145/3658644.3690354} embedding techniques due to their effectiveness in capturing complex relationships between entities. \ac{kg} is effective in representing entities and their interrelations through triples of the form $(h, r, t)$, where $h$ and $t$ are entities, and $r$ is the relation between them.
Existing \ac{kg} embedding methods~\cite{bordes2013translating, wang2014knowledge, lin2015learning, sun2019rotate, yang2015embedding} learn the embedding for entities and relations that satisfy $e_{h} + e_{r} \approx e_{t}$, where $e_{h}$, $e_{r}$, and $e_{t}$ are vectors in high dimensions. 
In our scenario, the relation between users in the dimension of $r_{1}$ can be represented as two triple $(u_{i}, r_{1}, u_{j})$ and $(u_{j}, r_{1}, u_{i})$, where $u_{i}$ and $u_{j}$ represent users and $r_{1}$ represents the relation type. We represent this relation between user $u_{i}$ and user $u_{j}$ in two triples because the relation between users is symmetric in the dimension of $r_{1}$.
In this way, the \ac{kg} can represent the relationship between users in multiple dimensions. TransR~\cite{lin2015learning} is particularly suitable for our needs as it can handle 1-to-N, N-to-1, and N-to-N relationships, which are common in user relationships in e-commerce platforms.

Formally, TransR embeds user entities $u_{i}, u_{j}$ and relation $r$ into vectors $\mathbf{e}_{u_{i}}, \mathbf{e}_{u_{j}} \in \mathbb{R}^{d}, \mathbf{e}_{r} \in \mathbb{R}^{k}$. For each relation $r$, it specifies a projection matrix $\mathbf{W}_{r} \in \mathbb{R}^{d \times k}$ to project the user entities from $d$-dimensional space to $k$-dimensional space. The embedding of the relation $r$ is calculated as $\mathbf{e}^{r}_{u_{i}} = \mathbf{e}_{u_{i}}\mathbf{W}_{r}$. The score function of TransR model is defined as $f_{r}(i,j) = ||\mathbf{e}_{u_{i}}^{r} + \mathbf{e}_{r} - \mathbf{e}_{u_{j}}^{r}||$, where $||\cdot||$ denotes the L1-norm distance function. To optimize the TransR model, we follow the margin-based score function as the objective for training
\begin{equation}\label{eqa:transr}
\mathcal{L}_{r} = \sum_{(i,r,j)\in S} \sum_{(i',r,j')\in S'} \max(0,f_{r}(i,j) + \gamma - f_{r}(i',j'))
\end{equation}
, where $S$ and $S'$ are the positive and negative triple sets, respectively, and $\gamma$ is the margin hyperparameter.
However, the embedding learning of TransR cannot deal with the new entities as it can only learn the embedding of the entities that appear in the training set. 
While in our scenario, the new users may appear in the production environment.
To address this challenge, we only use the embeddings of the relations $E_{r} = \{\mathbf{e}_{r_{1}}, \mathbf{e}_{r_{2}}, ..., \mathbf{e}_{r_{|\mathcal R|}}\}$ to initialize the relation embedding for later training. 

\noindent \textbf{Weighted Concatenation.}
After getting the embeddings of each type of relation, we propose a weighted concatenation method to get the final embedding of the relation between users.  
Although the high cohesion in spatial and temporal is the fundamental characteristic of fraudulent transactions, users can show different cohesion in different dimensions. For example, the users can show high cohesion in the dimension of the geohash and the dimension of the promotion, but they may show low cohesion in the dimension of the chat room. To better capture the divergence of the cohesion in different dimensions, we propose concatenating the embeddings of the relations into one vector. We utilize the concatenation operation because it can better capture the cohesion in multiple dimensions compared with the mean operation. A naive approach is to concatenate the embeddings of each type of relation if there is a relation between users, otherwise set the embedding of the relation to zero. However, this approach cannot capture the strength of the relationship between users in different dimensions. Existing \ac{kg} embedding methods treat all the relations equally and do not consider the strength of each relation.

To consider the strength of the relations between users, we propose to concatenate the relation embeddings with the weighted concatenation to get the final embeddings $\mathcal F$ of the relations between users. 
As we have discussed in Section~\ref{sec:study:analysis}, the co-occurrence frequency of users in different spatial relations, denoted as $w_{r}^{ij}$, effectively quantifies the strength of temporal cohesion of users. This metric captures the strength of relationships across various dimensions. Fraudsters are typically organized in groups and exhibit a higher value of this metric in spatial relations compared to normal users.
Therefore we can utilize $w_{r}^{ij}$ to quantify the strength of the relations between users. Therefore, the final embeddings of the relations between users are calculated as follows:

\begin{equation}
    \mathbf{f}_{ij} = \mathbin\Vert_{r = 1}^{|\mathcal R|}
    \begin{cases}
        \hfil w_{r}^{ij} \cdot \mathbf{e_{r_{i}}}&, m_{ij}(r) = 1 \\
        \hfil \mathbf{0} & , m_{ij}(r) = 0 \\
    \end{cases}
\end{equation}
, where $m_{ij}$ is the relation map and $m_{ij}(r)$ represents whether there is relation $r$ between users $u_{i}$ and $u_{j}$, $\mathbin\Vert$ is the concatenation operation. If there is no relation between users $u_{i}$ and $u_{j}$ in the dimension $r$, we set the embedding of the relation as vector $\mathbf{0}$, which is the same size as the embedding of the relation. Then we get the final embeddings of the relations between users $\mathcal F = \{\mathbf{f}_{ij}\}$ considering both the spatial and temporal information.

\subsection{Group-based Fraud Detection}
Having obtained the fused embedding relations between users, the objective is to detect fraudsters involved in promotion abuse on the multi-relation fused homogeneous graph. 
Fraudsters conduct fraudulent transactions in groups and exhibit high cohesion across multiple dimensions, which is reflected in the embeddings of the relations between users. Consequently, they form more condensed clusters in various dimensions on the graph compared to normal users.
To discover this condensed characteristic, we propose a fraud detection algorithm based on a multi-relation fused homogeneous graph. The detection workflow comprises three main modules: attention-based feature aggregation for learning user representations, a semi-supervised autoencoder for detecting fraudster seeds and anomaly propagation for identifying fraudulent neighbors in the same group. 

\noindent\textbf{Attention-based Node Feature Aggregation.} 
This module takes the multi-relation fused homogeneous graph $\mathcal{G}$ as input. Its goal is to compute user node features $\mathbf{h}_i'$ based on fused relation embeddings, enabling clear distinctions between fraudulent and normal users. Fraudsters engaged in promotion abuse fraud often operate in groups and exhibit high cohesion across multiple dimensions. To capture this behavior, we employ a \ac{gnn} to aggregate features from neighboring nodes $\mathcal{N}(i)$.
Since fraudsters are not always malicious, the influence of neighbors on a node's behavior varies. To address this, we use an attention mechanism to assign adaptive weights to neighbors, prioritizing those showing strong cohesion across dimensions. To better detect fraudulent behaviors regardless of promotion strategy, user node features are initialized as uniform vectors.
The attention score $\alpha_{ij}$ between user $i$ and user $j$ is calculated based on the node feature $\mathbf{h}_i$ and their edge feature $\mathbf{f}_{ij}$. Node features are then aggregated using these attention scores, as follows:

\begin{equation}
    \alpha_{ij} = \frac{1}{l} \sum_{k = 1}^{l} \text{LeakyReLU}(\mathbf{h}_{i}\mathbf{W}_{k}(\mathbf{f}_{ij}\mathbf{W})^{T})
\end{equation}

\begin{equation}
    \mathbf{h}_i' = \sum_{j \in \mathcal{N}(i)} \alpha_{ij} \mathbf{h}_j
\end{equation}

, where $\mathbf{W}_{k} \in \mathbb{R}^{D_{n} \times D_{a}}$ and $\mathbf{W} \in \mathbb{R}^{D_{e} \times D_{a}}$ are weight matrices, $D_{a}$ represents the attention size, and $l$ is the number of attention layers.

\noindent\textbf{Semi-supervised AutoEncoder.}
The aggregated node features $\mathbf{h}_i'$ are input to a semi-supervised autoencoder designed to identify fraudulent users.
In real-world scenarios, user labels are often limited, with only a small proportion of users accurately labeled through manual inspection. Additionally, fraudsters may change their behaviors in response to changes in promotion strategies. To address these challenges, we use a semi-supervised autoencoder to guide training and maintain the model's generalization based on the fact that fraudsters are rare and most users are normal.
The autoencoder consists of an encoder—a multi-layer perceptron mapping node features to a latent space—and a decoder that reconstructs node features $\mathbf{h}_i^*$ from the latent space. It is trained to minimize the reconstruction loss $l = ||\mathbf{h}_i' - \mathbf{h}_i^{*}||_{2}$.

The model's total loss comprises three components: the relation embedding loss $\mathcal{L}_{r}$ as defined in Equation~\ref{eqa:transr}, the reconstruction loss of the labeled users $\mathcal{L}_{l} = \mathrm{BCELoss}_{y_{i} \in \{0, 1\}}(y_{i}, l_{i})$ and the unlabeled users $\mathcal{L}_{u} = \mathrm{MSELoss}_{y_{i} = -1}(l_{i})$. The $\mathrm{BCELoss}$ is the binary cross-entropy (BCE) loss function and is defined as $\mathrm{BCELoss}(y, l) = -[ylogl + (1 - y)log(1-l)]$. The $\mathrm{MSELoss}$ is the mean squared error (MSE) loss function. 
The total loss function is defined as:
\begin{equation}
    \mathcal{L} = \mathcal{L}_{r} + \mathcal{L}_{l} + \mathcal{L}_{u}
\end{equation}

The model is trained offline using transaction data from the most recent $T$ days (default $T=7$). Fraudsters are detected based on reconstruction losses, with the top $T_s=1.2\%$ reconstruction losses designated as fraudster seeds. The setting of $T_s$ is based on the observation that fraudsters are around 1.5\% in the production environment and we evaluate the performance of the model with different $T_s$ in Section~\ref{sec:rq4}.

\noindent \textbf{Anomaly Propagation.} Although reconstruction-based anomaly detection is more generalizable than supervised methods, balancing precision and recall remains a challenge~\cite{10144292}. 
Moreover, fraudsters within the same group may exhibit varying reconstruction losses due to differences in their neighbors, resulting in potential false negatives. 
To address this limitation, we propose using fraudsters with high reconstruction errors as anomaly seeds and propagating anomalies on a multi-relation fused graph to identify additional fraudsters. 

To enhance detection, we extend the fraudster seed set using an anomaly propagation strategy. Fraudsters typically exhibit strong cohesion across multiple dimensions. We quantify this cohesion using the attention score $\alpha_{ij}$ and the weight of each relation $w^{ij}_{r}$. 
The propagation score between user $i$ and user $j$ is defined as:
\begin{equation}
    p_{ij} = \sum_{r = 1}^{|\mathcal R|} \alpha_{ij}*w^{ij}_{r}
\end{equation}
Using the propagation score, we identify fraudsters who exhibit high cohesion across multiple dimensions with the seed fraudsters. This propagation strategy computes scores in single pass using precomputed attention scores, enabling quick updates on \Company's production environment.
For each fraudster seed $i$, we compute $p_{ij}$; if $p_{ij}$ exceeds a threshold $T_p$, the user $j$ is classified as a fraudster within the same group. We also evaluate the performance of the model with different $T_p$ in Section~\ref{sec:rq4}.

%% file: tex/eval.tex
\section{Evaluation}\label{sec:evaluation}
We evaluate the performance of \toolname in detecting promotion abuse fraudsters in e-commerce platforms of \Company. We compare \toolname with five group-based fraud detection baselines: GraphSAGE~\cite{10.5555/3294771.3294869}, FRAUDAR~\cite{hooi2016fraudar}, GFDN~\cite{yu2023group}, COFRAUD~\cite{zang2023don}, and DiG-In-GNN~\cite{zhang2024dig}. We evaluate the performance of \toolname and the baseline methods on a publicly accessible dataset \ac{ppa}. We also deploy \toolname and baselines in the production environment of \Company. To evaluate the effectiveness of \toolname, we propose the following five research questions:

\begin{itemize}[noitemsep, topsep=1pt, partopsep=1pt, listparindent=\parindent, leftmargin=*]
    \item \textbf{RQ1:} Can \toolname effectively detect promotion abuse fraudsters in e-commerce platforms?
    \item \textbf{RQ2:} What are the main causes of false positives and false negatives in \toolname's detection of promotion abuse fraudsters?  
    \item \textbf{RQ3:} How does \toolname perform in real-world deployment?
    \item \textbf{RQ4:} How does each part of \toolname contribute to the detection performance?
    \item \textbf{RQ5:} How do hyperparameters affect the detection performance of \toolname?
\end{itemize}

\subsection{Experiment Setup}
In this section, we introduce the experiment setup, including the dataset usage, baselines, and the implementation and deployment of \toolname.

\noindent \textbf{Dataset Usage.} 
We evaluate the performance of \toolname and the baseline methods using datasets from \Company. For publicly accessible evaluation, we evaluate on the \ac{ppa}. The statistics of \ac{ppa} are shown in Table~\ref{tab:publicdataset}. The first week of the dataset is used for training and validation, while the second week is reserved for testing. The dataset and documentation are publicly available at \url{https://osf.io/rasje/}.
For large-scale evaluation, we deployed \toolname in the production environment of \Company and collected the statistics for one week. For each day, we detect the fraudsters using the transaction data from the previous week.

\noindent \textbf{Baseline.} We select five graph-based fraud detection methods as baselines, including FRAUDAR~\cite{hooi2016fraudar}, GFDN~\cite{yu2023group}, COFRAUD~\cite{zang2023don} and DiG-In-GNN~\cite{zhang2024dig}, GraphSAGE~\cite{10.5555/3294771.3294869}. These methods include state-of-the-art graph-based models for fraud detection and traditional \ac{gnn} models.

For GraphSAGE and DiG-In-GNN, they require node features as input. Therefore, we initialize the node features as a 12-dimensional feature vector about the user's transaction behavior during the recent week, including the number of transactions, the kinds of products, the number of products, etc. 
For GraphSAGE, we merge the multi-relation graph into a homogeneous graph to learn the node embeddings and use the same semi-supervised autoencoder to detect the fraudsters. For DiG-In-GNN, it only supports three types of relations. Therefore, we only consider the top three relations with the highest proportions, including $r_4$, $r_6$ and $r_8$. For FRAUDAR, GFDN and COFRAUD, we use the open-source implementation provided by the authors. Due to the large scale of the dataset, we set the $\alpha_{\tau}^{+} = \alpha_{\tau}^{-} = 4, \beta_{\tau}^{-} = 1, \beta_{\tau}^{+} = 100$ in GFDN to ensure the computation can be completed within a limited time. The other hyperparameters are set to the default values provided by the original implementation. 

\noindent \textbf{Implementation and Deployment.} We implement \toolname in Python and train our model using PyTorch~\cite{paszke2019pytorch}. For \ac{kg}-based relation embedding, we use the TransR model to learn the relation embeddings. The dimension of the relation embeddings $d$ is set to 8. Then we use the learned relation embeddings to initialize the relation embeddings in the \toolname. The dimension of the concatenated relation embeddings $D_{e}$ is set to 64. In graph-based fraud detection, we initialize the node embeddings of our model using full one vectors and project them to $D_{n} = 52$ with a linear layer. The attention size $D_{a}$ is set to 8, and the number of attention layers $l$ is set to 3.
We split the training data with a 5:2 ratio between training and validation sets. During training, we monitor the F1-score on the validation set every 100 epochs and implement early stopping if no improvement is observed for 5 consecutive checks. Training typically converges within 2,000 epochs, which we set as the default value. We use the Adam optimizer with a learning rate of 0.0001.
We set $T_s = 1.2\%$ to consider the top $T_{s}$ reconstruction losses as fraudster seeds. For anomaly proportion, we set $T_p = 0.65$ to identify fraudsters within the same group. We evaluate different threshold values of $T_s$ and $T_p$ to analyze the impact of these hyperparameters on the detection performance in Section~\ref{sec:rq4}. We deploy \toolname in the clusters of \Company with 96GB memory and an NVIDIA Tesla A100 GPU.
Our implementation is publicly available at \url{https://github.com/0xllssFF/PromoGuardian}.

\subsection{RQ1: Detection Performance}
\label{sec:rq1}

\begin{table}[t!]
    \scriptsize
    \centering
    \caption{Detection performance of \toolname and baselines on \ac{ppa}. \toolname-R/W/P represents the ablation study of \toolname.}
    \resizebox{0.49\textwidth}{!}{\begin{tblr}{
        cell{1}{1-6} = {c},
        cell{2-11}{2-6} = {r},
        vline{2},
        hline{1,11} = {1pt},
        stretch=0
    }
    \SetCell[r = 1, c = 1]{c} \textbf{Method} & \SetCell[r = 1, c = 1]{c} \textbf{Precision} & \SetCell[r = 1, c = 1]{c} \textbf{Recall} & \SetCell[r = 1, c = 1]{c} \textbf{F1} & \SetCell[r = 1, c = 1]{c} \textbf{Accuracy} \\ \hline
    \SetCell[r = 1, c = 1]{c} \textbf{FRAUDAR} & 0.4667 & 0.4765 & 0.4715 & 0.9846  \\
    \SetCell[r = 1, c = 1]{c} \textbf{GFDN} & 0.3225 & 0.4314 & 0.3690 & 0.9018 \\
    \SetCell[r = 1, c = 1]{c} \textbf{COFRAUD} & 0.1386 & 0.1855 & 0.1587 & 0.9596 \\
    \SetCell[r = 1, c = 1]{c} \textbf{DiG-In-GNN} & 0.1729 & 0.1650 & 0.1689 & 0.9607  \\
    \SetCell[r = 1, c = 1]{c} \textbf{GraphSAGE} & 0.6110 & 0.1824 & 0.2810 & 0.9564\\ \hline
    \SetCell[r = 1, c = 1]{c} \textbf{\toolname-R} & 0.8383 & 0.4122 & 0.5527 &  0.9829 \\
    \SetCell[r = 1, c = 1]{c} \textbf{\toolname-W} & 0.7332 & 0.3890 & 0.5083 & 0.9805 \\
    \SetCell[r = 1, c = 1]{c} \textbf{\toolname-P} & 0.8546 & 0.5382 & 0.6605 & 0.9912 &\\ \hline
    \SetCell[r = 1, c = 1]{c} \textbf{\toolname} &  \textbf{0.9107} & \textbf{0.6992} & \textbf{0.7911} & \textbf{0.9923} \\

\end{tblr} }
    \label{tab:publicresult}
\end{table}

We evaluate the detection performance of \toolname and baseline methods on \ac{ppa} to demonstrate its effectiveness in identifying promotion abuse fraudsters. The results, presented in Table~\ref{tab:publicresult}, are measured using precision, recall, F1 score, and accuracy. For the detected unlabeled data, the findings were reported to the policy team of \Company. They randomly sampled 100 groups (where users purchasing products from the same retail store are considered one group) from the detected unlabeled data and manually verified the user labels within these groups. They group the users based on the retail store where they placed the orders, as the retail store is a key factor in ordering and delivery. Therefore the fraudsters always carry out fraudulent activities under the same retail store. 
They calculated the true positives and false positives for the sampled users and extrapolated the statistics to all unlabeled users based on the sampling proportion. The overall metrics were then computed by combining these statistics with those of the labeled users. Finally, precision, recall, F1 score, and accuracy were derived from the aggregated statistics.

\begin{table}[t!]
    \scriptsize
    \centering
    \caption{The number of detected users of \toolname and baselines on \ac{ppa}. TP denotes true positive, FP denotes false positive. L and U denote labeled and unlabeled users. \toolname-R/W/P represents the ablation study of \toolname.}
    \resizebox{0.4\textwidth}{!}{\begin{tblr}{
        cell{1}{1-6} = {c},
        cell{2-12}{2-6} = {r},
        vline{2},
        hline{1,12} = {1pt},
        stretch=0
    }
    \SetCell[r = 2, c = 1]{c} \textbf{Method} & \SetCell[r = 1, c = 2]{c} \textbf{TP} & & \SetCell[r = 1, c = 2]{c} \textbf{FP} &  \\ \cline{2-9}
    & \SetCell[r = 1, c = 1]{c} \textbf{L} & \SetCell[r = 1, c = 1]{c} \textbf{U} & \SetCell[r = 1, c = 1]{c} \textbf{L} & \SetCell[r = 1, c = 1]{c} \textbf{U} \\ \hline
    \SetCell[r = 1, c = 1]{c} \textbf{FRAUDAR} & 43,129 & 15,469 & 683 & 66,277  \\
    \SetCell[r = 1, c = 1]{c} \textbf{GFDN} &38,517  & 14,535 & 1,862 & 109,588 \\
    \SetCell[r = 1, c = 1]{c} \textbf{COFRAUD} & 17,246 & 5,566 & 2,535 & 139,242   \\
    \SetCell[r = 1, c = 1]{c} \textbf{DiG-In-GNN} & 16,152 & 4,139 & 1,263 & 95,803  \\
    \SetCell[r = 1, c = 1]{c} \textbf{GraphSAGE} & 17,407 & 5,024 & 451 & 13,830  \\ \hline
    \SetCell[r = 1, c = 1]{c} \textbf{\toolname-R} & 36,803 & 13,888 & 402 & 9,376  \\
    \SetCell[r = 1, c = 1]{c} \textbf{\toolname-W} & 35,688 & 12,150 & 534 & 16,873  \\
    \SetCell[r = 1, c = 1]{c} \textbf{\toolname-P} & 54,671 & 11,515 & 420 & 10,814   \\ \hline
    \SetCell[r = 1, c = 1]{c} \textbf{\toolname} & 65,006 & 20,979 & 356 & 8,075  \\

\end{tblr} }
    \label{tab:statistics}
\end{table}

The results show that \toolname outperforms all baseline methods in terms of precision, recall, F1 score, and accuracy. Specifically, \toolname achieves a precision of 0.9107, a recall of 0.6992, an F1 score of 0.7911, and an accuracy of 0.9923. 
All the baseline methods have high accuracy because the normal users are the majority in the dataset, which results in a high true negative rate.
FRAUDAR and GFDN achieve the best performance among the baseline methods because promotion abuse in the same group always shows aggregation on the limited products, which can be detected by the dense subgraph detection method. However, they only achieve a precision of 0.4667 and 0.3225 because they only consider the dense subgraph detection and ignore the spatial and temporal relations between transactions, which results in false positives on popular products and legitimate dealers. COFRAUD leverages the alienation and marginalization characteristics of fake reviews to detect fraudsters, which is not proper in detecting promotion abuse fraudsters. DiG-In-GNN only supports limited relations and loses lots of relation information on the graph, which results in lower performance. GraphSAGE achieves the best precision among the baseline methods, but it has a lower recall. 

Table~\ref{tab:statistics} provides detailed manually verified results on the number of true positives and false positives for \toolname and the baseline methods. \toolname detects 65,006 labeled fraudsters and 20,979 unlabeled fraudsters, with only 8,431 false positives. 
To further understand \toolname's effectiveness, we apply two rules from Section~\ref{sec:study:company} to classify the types of detected fraudulent groups for FRAUDAR, the best among the baseline methods, and \toolname. FRAUDAR identifies 2,945 groups of fraudsters, with 88\% of the detected groups classified as stocking-up fraud, which typically involves purchasing popular products in large quantities. Additionally, only 16\% of the groups are identified as cashback abuse fraud, most of which overlap with stocking-up fraud. 
In contrast, \toolname detects 4,150 groups of fraudsters, with 45\% classified as cashback abuse fraud and 68\% as stocking-up fraud. This demonstrates that \toolname is more effective at detecting cashback abuse fraud compared to FRAUDAR. FRAUDAR primarily focuses on dense subgraph detection, which is effective for identifying stocking-up fraud. However, \toolname leverages spatial and temporal cohesion patterns in fraudulent activities, enabling it to detect fraud that may not exhibit high density but still demonstrates significant cohesion across spatial and temporal dimensions.

\subsection{RQ2: False Positives and False Negatives}

We analyze its false positives and false negatives in identifying promotion abuse fraudsters. Given the large scale of the dataset, manually verifying all cases is infeasible. Therefore, we first group users based on the retail store from which they place the orders, as described in Section~\ref{sec:rq1}. We then manually examine the top ten largest groups of users to identify the main causes of them.

For false positives, we observe that all ten retail stores associated with these groups exhibit fraudulent activities during our detection period. This indicates that the users in these groups are not fraudsters themselves but happen to purchase products from the same retail store. Most of these false positives occur because these users either made purchases during the same time period or ordered a small number of products from the same retail store within the detection window, creating patterns that resemble fraudulent behavior. Specifically, 92\% of these users purchased high-circulation products that the fraudsters targeted for and 79\% had less than five transactions during the detection period, which makes them look similar to fraudsters. They are not labeled as fraudsters because they do not exhibit significant fraudulent behavior in terms of purchase amount and frequency.
Therefore, these users are not penalized by the platform or blocked from making purchases. The platform assesses the risk of detected users based on manually crafted rules, primarily considering the frequency of purchases and the amount of money spent. Since these users do not exhibit significant fraudulent behavior in terms of purchase amount and frequency, they are not classified as high-risk users. We then feed all the false positives to the rules and only 479 users are classified as medium-risk users, which is acceptable for the platform.

For false negatives, we find that 85\% of them are caused by fraudsters operating in small groups of fewer than ten members. 
The largest group among these false negatives contains 26 members who split into smaller subgroups to purchase products from five different retail stores. During the detection period, these subgroups were not fixed and the combination of members changed frequently. As a result, the group exhibited less cohesion on the graph, making detection more challenging.
The false negatives are labeled as fraudsters because the label is based on the reported cases and existing XGBoost-based classification methods. 
Existing XGBoost-based fraud detection methods classify them as fraudsters based on the statistics of their transaction features, such as the number of products purchased in the past 7, 14, and 30 days. Therefore, although these fraudsters are in small groups, they can be detected by the existing XGBoost-based methods. However, they do not exhibit significant cohesion in the graph, making them harder to detect using \toolname. 
Compared to the large-scale fraud, these small groups typically do not cause substantial economic losses to the platform. Consequently, \toolname is complementary to the existing XGBoost-based approach and can enhance the overall protection level of the platform.

\subsection{RQ3: Real-world Deployment}
To demonstrate the effectiveness and scalability of \toolname in real-world operational conditions, we deploy \toolname in the production environment of \Company and evaluate its performance for one week in November 2024. In real-world production environments, the daily active users of \Company reach millions, and the daily transaction numbers are in the tens of millions. We also compare the detection performance of \toolname with the baseline methods. 

We deploy the model trained on \ac{ppa} to detect fraudsters in the production environment. 
For online deployment, we cannot calculate the recall, F1 score, and accuracy because the ground truth is not available. Therefore, we only evaluate the precision and actual effect of the detection, which are more important in real-world scenarios.
For precision and the number of detected fraudsters, we use the manual rules for risk grading that are deployed in the online system and regard all risky users as fraudsters. We then calculated the number of transactions blocked by the system and the economic losses avoided by blocking these transactions. The Gross Margin (GM) for each transaction is a key indicator for the platform to evaluate the economic income, which is defined as the sales revenue minus the cost of sales. 
For most of the products under promotion, the subsidies provided by the platform exceed the actual revenue generated, resulting in negative GM. 
Therefore the avoided economic losses can be calculated as: $L(T) = -\sum_{i\in T}{GM_{i}}$, where $T$ is the set of transactions blocked by the system and $GM_{i}$ is the GM of transaction $i$.

The average results of the seven days are shown in Table~\ref{tab:productionresult}.
The results show that \toolname achieves the best detection performance among all the methods, with a precision of 0.9315, 37,517 fraudsters, 72,734 blocked transactions, and \$27,945 avoided economic losses per day. 
Among the detected fraudsters, 76\% of them are ordinary users who also purchase products that are not under promotion during the detection period, which indicates that these users conducted transactions for their own purposes or to camouflage their behaviors. The results demonstrate that \toolname can effectively detect promotion abuse in the production environment and avoid economic losses.

\begin{table}[t!]
    \scriptsize
    \centering
    \caption{The average detection performance in the production environment for a week. \toolname-R/W/P represents the ablation study of \toolname.}
    \resizebox{0.48\textwidth}{!}{\begin{tblr}{
        cell{3-11}{2-5} = {r},
        hline{1,12} = {1pt},
        stretch=0
    }
    \SetCell[r = 2, c = 1]{c} \textbf{Method} & \SetCell[r = 2, c = 1]{c} \textbf{Precision}  & \SetCell[r = 1, c = 1]{c}  \textbf{Detected}  &  \SetCell[r = 1, c = 1]{c}  \textbf{Blocked} & \SetCell[r = 1, c = 1]{c} \textbf{Avoided Economic} \\ 
    & & \SetCell[r = 1, c = 1]{c} \textbf{Fraudsters} & \SetCell[r = 1, c = 1]{c}  \textbf{Transactions} & \SetCell[r = 1, c = 1]{c}  \textbf{Losses (\$)} \\ \hline
    \SetCell[r = 1, c = 1]{c} \textbf{FRAUDAR} & 0.4207 & 13,628 & 58,243 & 14,515 \\
    \SetCell[r = 1, c = 1]{c} \textbf{GFDN}  & 0.2859 & 17,583 & 59,174 & 17,534\\
    \SetCell[r = 1, c = 1]{c} \textbf{COFRAUD}  & 0.1714 & 7,527 & 15,254 & 5,226 \\
    \SetCell[r = 1, c = 1]{c} \textbf{DiG-In-GNN} & 0.1627 & 8,911 & 10,969 & 3,159  \\
    \SetCell[r = 1, c = 1]{c} \textbf{GraphSAGE} & 0.4600 & 15,444  & 30,899 & 10,044 \\ \hline
    \SetCell[r = 1, c = 1]{c} \textbf{\toolname-R} & 0.7251 & 20,509 & 54,569 & 19,035 \\
    \SetCell[r = 1, c = 1]{c} \textbf{\toolname-W} & 0.5987 & 17,583  & 52,304 & 16,603 \\
    \SetCell[r = 1, c = 1]{c} \textbf{\toolname-P} & 0.6981 & 28,326 & 62,680 & 21,537  \\ \hline
    \SetCell[r = 1, c = 1]{c} \textbf{\toolname} & \textbf{0.9315} & \textbf{37,517} & \textbf{72,734} & \textbf{27,945} \\

\end{tblr} }
    \label{tab:productionresult}
\end{table}

\begin{figure}[t!]
    \centering
    \includegraphics[width=0.49\textwidth]{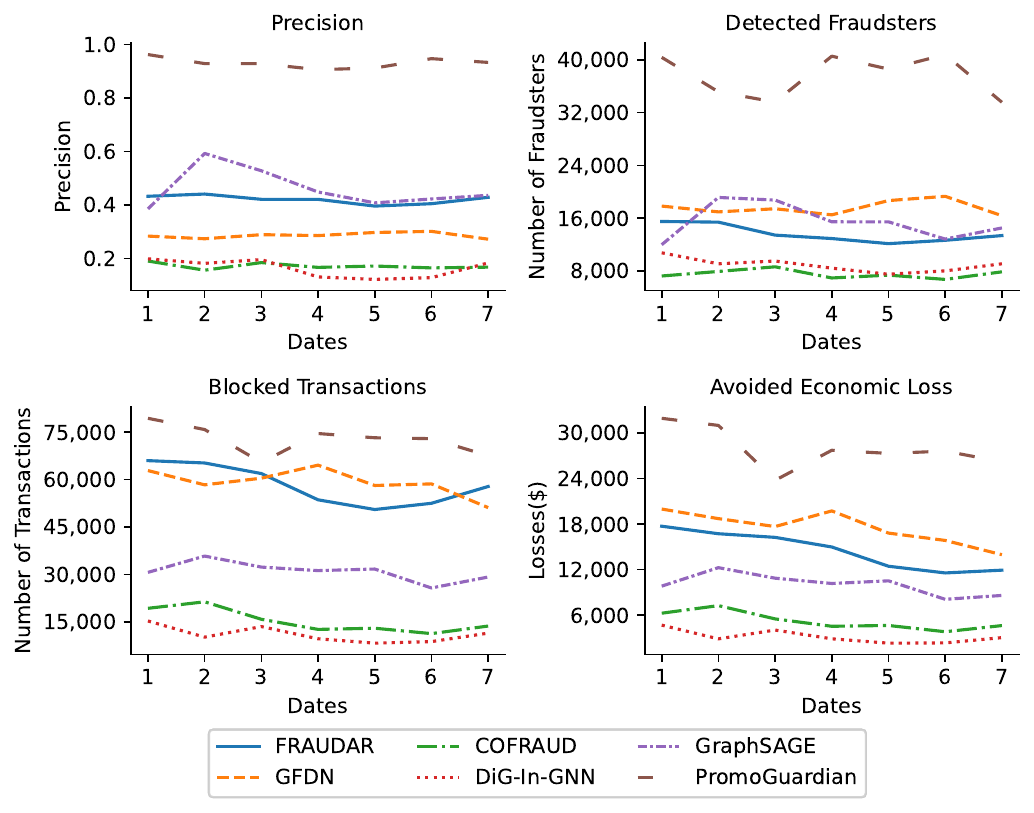}
    \caption{The detection performance of \toolname and baselines in real-world deployment.}
    \label{fig:productionresult}
    
\end{figure}

Figure~\ref{fig:productionresult} shows the results of each day and reflects the stability of \toolname in the production environment. The results show that \toolname consistently outperforms the baseline methods in our evaluation. 
Among the baseline methods, FRAUDAR and GFDN blocked more transactions and avoided more economic losses than other methods, while detecting similar numbers of fraudsters as COFRAUD and GraphSAGE. This is because FRAUDAR and GFDN are more sensitive to the dense subgraphs in the graph, therefore they are effective in detecting fraudsters in large groups with more transactions. 
However, they also have a higher false positive rate, which results in lower precision. These false positives are mainly caused by legitimate dealers who have a large number of transactions and are not considered fraudsters. Therefore, we need to consider the cohesion of the fraudsters in spatial and temporal relations to reduce the false positives in such cases.

\noindent \textbf{Case Study.} During the one-week deployment of \toolname, a large-scale promotion abuse fraud group was uncovered. This group involved 319 users, with three headers (Header X, Header Y, and Header Z) acting as the organizers. The retail stores of these three headers are located in the same city, and they are responsible for managing the sales and deliveries for different regions of the city. The group engaged in stocking-up fraud over two days and cashback abuse fraud for an entire week.

In the case of stocking up, they targeted high-demand products with good resale value. During the evaluation period, a promotion on drinks such as milk, beer, and water offered subsidies, making the products cheaper than on other platforms. Each user was limited to purchasing two packs. To bypass this limit, the headers organized multiple accounts. For instance, Header X coordinated with 96 users, who collectively placed 183 orders of drinks within a week, far exceeding normal purchasing volumes. These purchases were often made within a 30-minute window on specific days, with orders placed at the same location or using the same shared link. Similar patterns were observed in the retail stores of Header Y and Header Z.

For cashback abuse, the platform provides cashback incentives on specific products it aims to promote. The three headers colluded to fabricate transactions and claim cashback rewards. They shared their user networks and organized users to purchase low-cost products, such as instant noodles, toilet paper, and snacks. For instance, Header Y's store in the eastern part of the city received 351 orders in a week from users associated with Header X in the north and Header Z in the west. Each user purchased only one of these low-cost products per day, and this pattern persisted for weeks. The same group of users consistently appeared during specific time periods, exhibiting high co-occurrence frequency.

Overall, this group placed 2,565 orders, with a total transaction value of \$13,354 and $L(T) =$ \$1,023.
The risk grading system of \Company identified 162 users as high-risk, banning their accounts for half a year; 101 users as medium-risk, banning their accounts for one month; and 56 users as low-risk, issuing warnings. The headers were penalized by suspending their promotion and cashback incentives for six months.

\subsection{RQ4: Ablation Study}
In this section, we evaluate the contribution of each component of \toolname through ablation studies. We evaluate three variants of \toolname, including \toolname-R, which removes \ac{kg}-based relation embeddings and initializes them with uniform vectors;
\toolname-W, which replaces weighted concatenation with simple concatenation by setting all relation weights to 1; and \toolname-P, which removes the anomaly propagation mechanism. 

\begin{figure}[t!]
    \centering
    \includegraphics[width=0.49\textwidth]{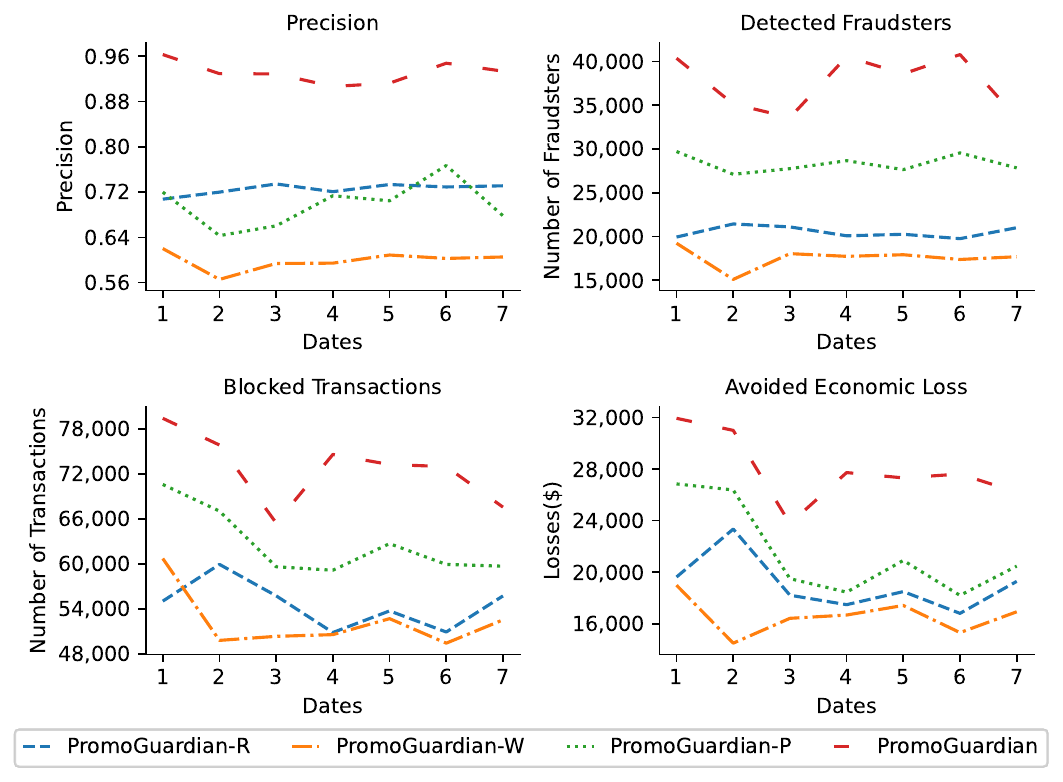}
    \caption{The ablation study of \toolname in real-world deployment.}
    \label{fig:ablation}
    
\end{figure}

We evaluate the performance of these three variants and compare them with the full model on \ac{ppa} and the production environment.
Table~\ref{tab:publicresult}, \ref{tab:statistics} and \ref{tab:productionresult} show the results on \ac{ppa} and the production environment. Figure~\ref{fig:ablation} shows the results of each day in the production environment.
Overall, the results show that all the variants of \toolname have decreased performance compared to the full model. Especially, \toolname-W performs the worst, with 35\% lower precision, 53\% lower detected fraudsters, 28\% lower blocked transactions, and 41\% lower avoided economic losses than the full model during the production environment evaluation. This is because the weighted concatenation can effectively capture the importance of each relation in the graph and reduce the noise caused by the irrelevant relations. \toolname-R and \toolname-P also have slightly decreased performance compared to the full model. \ac{kg}-based relation embedding can effectively capture the complex relations between users and get better relation embedding than initializing with full one vectors. Anomaly propagation can effectively propagate to discover the fraudsters in the same group.

\subsection{RQ5: Evaluation on Hyperparameters} \label{sec:rq4}
\begin{figure}[!t]
    \centering
    \includegraphics[width=0.48\textwidth]{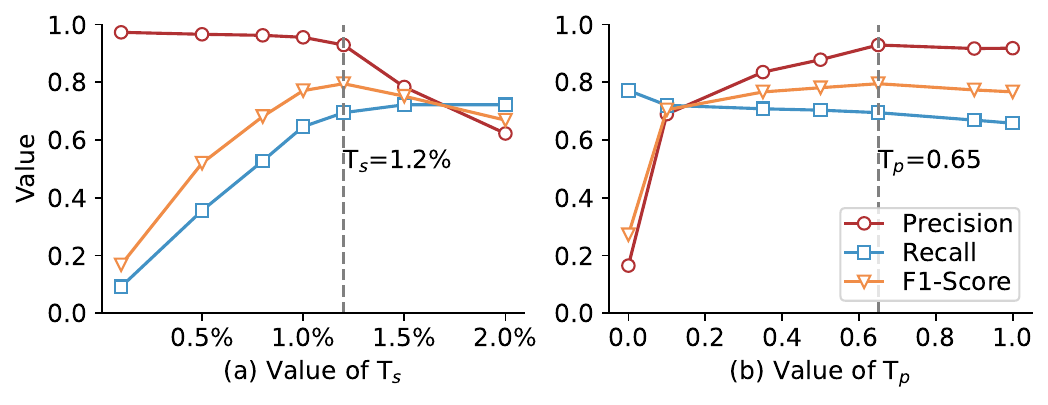}
    \caption{The impact of hyperparameters on the detection performance of \toolname.}
    \label{fig:hyperparameters}
\end{figure}

We evaluate the hyperparameter settings of \toolname to analyze their impact on detection performance. Specifically, we examine the effects of two thresholds in group-based fraud detection. The first threshold, $T_s$, considers the users with top $T_s$ reconstruction losses as fraudster seeds. The second threshold, $T_p$, propagates the anomaly to identify fraudsters within the same group. We assess the detection performance of \toolname with different values of $T_s$ and $T_p$ on \ac{ppa}. The results are shown in Figure~\ref{fig:hyperparameters}. We calculate the precision, recall, and F1-score of \toolname for various values. Accuracy is excluded due to insignificant differences among values, attributed to the large number of true negatives. We select the hyperparameters with the best F1-score performance as the default settings in our experiments.

\noindent \textbf{Reconstruction Loss Threshold $T_s$.} The results indicate that an increase in $T_s$ leads to a decrease in precision but an increase in recall. This occurs because fraudsters with higher reconstruction losses are more likely to be identified as fraudulent, making lower $T_s$ values yield higher precision. Conversely, to detect more fraudsters and improve recall, a higher $T_s$ is required. The F1-score is highest when $T_s = 1.2\%$. Therefore, we set $T_s = 1.2\%$ in our experiments.

\noindent \textbf{Anomaly Propagation Threshold $T_p$.}
The results demonstrate that as $T_p$ increases, precision improves while recall declines. A lower $T_p$ propagates anomalies to more users, enabling the detection of additional fraudsters. However, this also leads to a higher rate of false positives, reducing precision. Based on our experiments, we select $T_p = 0.65$ where \toolname achieves the highest F1-score.

%% file: tex/relatedwork.tex
\section{Related Work}

\noindent \textbf{Fraud Detection Based on Single Relation.}
Single-relation-based detection methods are widely used in various domains, including social networks~\cite{xu2021deep,10.1145/1866307.1866418,bhattacharya2023comprehensive,10.1145/3319535.3363198,rahman2012efficient}, e-commerce platforms~\cite{10.1145/3308560.3316586,liu2018geniepathgraphneuralnetworks,hooi2016fraudar,10291497}, and scam~\cite{ma2024careful,10.1145/3511808.3557454,10.1145/3627106.3627109,10.1145/3658644.3690234}. These methods typically model the problem as a single-relation graph, where each edge represents a specific type of relationship between nodes. This relation is always explicit and can be easily observed. For example, in social networks, the relation could be a friendship or follower relationship, while in e-commerce platforms, it could represent a transaction or interaction between users.
For instance, FRAUDAR~\cite{hooi2016fraudar} and GFDN~\cite{yu2023group} models transaction between users and products as bipartite graphs and performs fraud detection. However, single-relation-based detection methods often struggle to deeply explore the complex relationships between nodes. 

\noindent \textbf{Fraud Detection Based on Multiple Relations.}
Multi-relation-based fraud detection methods model the problem as a multi-relation graph to represent more complex relationships between nodes~\cite{10.1145/3292500.3330927,9679178,Liu_2020,Dou_2020,zhang2024dig,10.1145/3442381.3449989,zang2023don}. These methods leverage the rich information provided by multiple relations to improve detection performance. 
For example, GraphConsis~\cite{Liu_2020}, CARE-GNN~\cite{Dou_2020}, and DiG-In-GNN~\cite{zhang2024dig} focus on addressing graph inconsistency issues by filtering neighborhoods to detect camouflaged fraudsters. PC-GNN~\cite{10.1145/3442381.3449989} tackles the label imbalance problem in fraud detection through label and neighborhood sampling. However, these methods often rely on manually designed meta-paths to capture the relationships between nodes, which can be time-consuming and may not generalize well to different scenarios. Additionally, they fail to effectively utilize the spatial and temporal information of relations, which is critical for detecting promotion abuse fraud.

%% file: tex/discussion.tex
\section{Discussion} \label{sec:discussion}

\noindent \textbf{Adversarial Adaptation.}
Although \toolname effectively detects promotion abuse fraud by leveraging high cohesion in spatial-temporal, fraudsters may adapt their strategies to evade detection. For instance,
fraudsters may fragment groups to conceal the cohesion patterns or use synthetic/stolen identities to reduce transaction frequency per account. 
While such tactics can help evade detection, but this raises their operational costs. Through our investigation, complete avoidance of spatial-temporal cohesion remains challenging for large-scale coordinated fraud.

\noindent \textbf{Limitations.} We acknowledge several limitations of \toolname. First, \toolname relies on comprehensive spatial and temporal data. In cases where key relations are missing, detection performance may degrade to that of transaction statistics-based methods. Although our approach is generalizable to varying numbers of relations, missing critical data can limit effectiveness. Second, the current study focuses on stocking up and cashback abuse, the most prevalent fraud types on the platform. Other forms of promotion abuse may exist but are not covered in this work. Third, \toolname may not be directly applicable to other fraud types due to differences in user behavior. However, the underlying principles of graph construction and behavioral analysis are adaptable to other scenarios.

\noindent \textbf{Future Work.} One promising direction is to integrate more effective node features. Existing XGBoost-based methods utilize statistical features from transaction history, but simply integrating these features may not improve performance due to redundancy and influence from legitimate transactions. Designing discriminative node features remains an open challenge. Moreover, developing dynamic features that reflect temporal changes in user behavior could help identify evolving fraud strategies. Finally, investigating feature selection and efficient fusion methods to reduce redundancy and improve interpretability will be crucial for building robust and scalable fraud detection systems.

\noindent \textbf{Takeaways.} Our research highlights that the key to detecting group-based fraud lies in understanding fraudster behavior and characterizing user relationships. By analyzing these relationships, we can construct a user relationship graph, which is more critical than the design of detection model itself. Traditional methods overlook the intrinsic connections between transactions, whereas fraudsters in promotion abuse scenarios exhibit high cohesion in spatial and temporal dimensions. Building accurate and comprehensive user relation graphs significantly enhances detection performance.

\section{Ethics Considerations}\label{sec:ethics}
We carefully adhere to ethical guidelines related to human data research, including principles of informed consent, data anonymization, and user privacy protection. There is a detailed explanation of the methods we have adopted in addressing ethical issues related to real user data in our research and public datasets in this section.

\noindent \textbf{Ethical Review and Approval.}
Our research was conducted with IRB approval from \Company obtained before starting the study. The IRB includes legal experts well-versed in data protection laws and regulations, ethicists capable of providing in-depth ethical analysis, as well as external professionals for ensuring unbiased and thorough reviews. The IRB strictly adheres to ethical norms and the requirements of the General Data Protection Regulation (GDPR) and the Personal Information Protection Law (PIPL) for privacy management. It makes sure that all information is kept private, and prevents any unauthorized access, use or disclosure of user data. All data used in this study were reviewed and approved by the IRB at \Company.

\noindent \textbf{User Consent for Data Collection.}
The collection and use of user data in this study are governed by the \Company's publicly available Privacy Agreement. This agreement includes a clear and comprehensive consent form specifying: (1) the detailed categories of data collected, such as user identification information, network identification information, transaction log information and so on; (2) the purposes of data use, including statistical analysis, operational improvement, database construction, and the commercialization or academic research of de-identified information that cannot be traced to individual users. For example, location and preference data may be utilized to enhance products, services, or marketing strategies, as well as to support system improvements through technical upgrades, network maintenance, process development, and internal reporting; (3) the privacy protection measures implemented, including secure encryption and rigorous de-identification prior to machine learning, algorithm model training, or other data mining activities, to ensure the optimization of our product experience model.

Users provide consent to the platform's Privacy Agreement during registration and retain the right to withdraw their consent at any time. The agreement authorizes the use of personal information for statistical analysis and operational improvement, and permits the use of de-identified data for machine learning, database construction, and commercialization or academic research. All data analyses conducted in this study are fully compliant with the scope of this authorization, and no additional consent is required. This study relies on platform-level consent obtained via the standard Privacy Agreement, rather than study-specific consent for this particular research. While this approach is common for large-scale operational studies and justified by the impracticality of re-contacting all users for each analysis, it may limit the granularity of user awareness regarding specific research activities. We acknowledge this as a limitation and emphasize that all data use strictly adheres to the scope and protections defined in the platform's Privacy Agreement and IRB approval.

\noindent \textbf{De-identification of Personal Private Data.}
\Company strictly follows its user privacy policy when collecting and managing user information. For personal identifiers directly linked to users' real identities, such as real names, e-mail, and phone numbers, immediate de-identification and encryption measures are applied. Specifically, these identifiers are permanently replaced with tokens generated using the SHA-256 cryptographic hash function. The original sensitive data is securely stored in a centralized tokenization vault~\cite{Tokenization}. This process ensures that researchers and engineers, including those involved in this study, only have access to de-identified data and cannot retrieve the original information. For other data, \Company anonymized location information using Geohash to prevent disclosure of exact locations.

\noindent \textbf{Anonymization of Public Datasets.}
The anonymized public dataset is composed of two main components: transaction data and user relationship graph derived from the transaction data. To balance research utility and privacy protection, we implemented comprehensive anonymization strategies for both components.

For the transaction data, which contains 11 fields (including transaction time, user ID, product ID, geohash of the transaction's location, shared links, delivery information, retail store, group identification, promotion ID, coupon type, and sales strategies), all fields are anonymized to prevent information leakage. For transaction time, we anonymized it with precision degradation and random-time shift by removing time-of-day details and applying a time shift~\cite{anonym}. For the other fields, we applied the enumeration strategy~\cite{anonym}: every unique value in a column is replaced with a sequential number assigned upon its first occurrence, and subsequent occurrences use the same assigned number. This ensures that all original values are replaced with non-informative identifiers, making it impossible to reconstruct or infer the original information from the dataset. For example, user/product IDs are anonymized to protect individual identities while maintaining relational structure. For the user relationship graph, each entry represents an edge between two users and includes eight relation weights. All user nodes are replaced with the anonymized identifiers from the transaction data. The edges only indicate associations between users and do not contain transaction-specific details, further preventing inference of user behavior from edge attributes. Through these anonymization strategies, all user and transaction-related information cannot be inferred or reconstructed, effectively eliminating privacy risks.

While the anonymization methods described above substantially reduce the risk of re-identification, it is important to acknowledge that no anonymization technique can guarantee the complete elimination of such risks. All data sharing and publication are subject to IRB review and approval, ensuring that privacy risks are continually assessed and mitigated to the greatest extent possible.

%% file: tex/conclusion.tex
\section{Conclusion}
\label{sec:conclusion}
Promotion abuse fraud presents a unique challenge in e-commerce platforms. It involves ordinary customers conducting legitimate transactions, which makes it difficult to distinguish them from normal transactions. In this paper, we propose \toolname, a novel fraud detection system that leverages multi-relation fused homogeneous graph neural networks to detect promotion abuse fraud. Instead of focusing solely on the users and product information in the transactions, \toolname discovers and models the intrinsic connections between users through multiple relations involving spatial and temporal information. We evaluate \toolname in a real-world production environment in \Company. The results show that \toolname outperforms the state-of-the-art fraud detection methods in detection precision, the volumes of detected fraudsters, and the financial losses prevented. 

\section{Acknowledgment}
We thank the anonymous reviewers for their valuable comments. This work was partly supported by the National Key R\&D of China
(2022YFB4501802), Beijing Natural Science Foundation (L243010) and National Natural Science Foundation of China (62141208).